\documentclass[journal,twoside]{IEEEtran}
\ifCLASSINFOpdf
\else
\fi
\usepackage{amsmath}
\usepackage{amssymb}

\usepackage{amsfonts}
\usepackage[nice]{nicefrac}
\usepackage[pdftex]{graphicx}
\usepackage{array}
\usepackage[caption=false,font=footnotesize]{subfig}
\usepackage{bm}
\usepackage{stfloats}
\usepackage{color}
\usepackage{cite}
\usepackage{url}
\begin{document}
%
\title{MIMO Radar Waveform-Filter Design for Extended Target Detection from a View of Games}

%
%
%

\author{Zhou~Xu,
        Chongyi~Fan
        and~Xiaotao~Huang,~\IEEEmembership{Senior Member,~IEEE}
\thanks{Manuscript received *** **, 2018; accepted *** **, 2018. Date of publication *** **, 2018; date of current version *** **, 2018.\emph{(Corresponding author: Chongyi Fan.)}}
\thanks{Z. Xu is with the College of Electronic Science and
	Engineering, National University of Defense Technology, Changsha 410073, and also with the Colledge of Electronic Countermeasure,National University of Defense Technology, Hefei 230037,	China (e-mail: zhouzhou900521@126.com).}
\thanks{C. Fan and X. Huang are with the College of Electronic Science and Engineering,
	National University of Defense Technology, Changsha 410073, China(e-mail: chongyifan@nudt.edu.cn).}
\thanks{Color versions of one or more of the figures in this paper are available
	online at http://ieeexplore.ieee.org.}
\thanks{Digital Object Identifier **.****/TIP.2018.******.}}

%
%

\markboth{IEEE TRANSACTIONS ON SIGNAL PROCESSING,~Vol.~**, No.~**, **~2020}
{Xu \MakeLowercase{\textit{et al.}}: MIMO RADAR WAVEFORM-FILTER DESIGN FOR EXTENDED TARGET DETECTION FROM A VEIW OF GAMES}
%




\maketitle

\begin{abstract}
This paper studies the Two-Person Zero Sum(TPZS) game between a Multiple-Input Multiple-Output(MIMO) radar and an extended target with payoff function being the output Signal-to-Interference-pulse-Noise Ratio(SINR) at the radar receiver. The radar player wants to maximize SINR by adjusting its transmit waveform and receive filter. Conversely, the target player wants to minimize SINR by changing its Target Impulse Response(TIR) from a scaled sphere centered around a certain TIR. The interaction between them forms a Stackelberg game where the radar player acts as a leader. The Stackelberg equilibrium strategy of radar, namely robust or minimax waveform-filter pair, for three different cases are taken into consideration. In the first case, Energy Constraint(EC) on transmit waveform is introduced, where we theoretically prove that the Stackelberg equilibrium is also the Nash equilibrium of the game, and propose Algorithm 1 to solve the optimal waveform-filter pair through convex optimization. Note that the EC can't meet the demands of radar transmitter due to high Peak Average to power  Ratio(PAR) of the transmit waveform, thus Constant Modulus and Similarity Constraint(CM-SC) on waveform is considered in the second case, and Algorithm 2 is proposed to solve this problem, where we theoretically prove the existence of Nash equilibrium for its Semi-Definite Programming(SDP) relaxation form. And the optimal waveform-filter pair is solved by calculating the Nash equilibrium followed by the randomization schemes. In the third case, a more challenge task with Spectral Compatibility and Similarity Constraint(SC-SC) is considered. Algorithm 3 is devised to address this problem by leveraging on Majorization Minimization(MM) method, where the Stackelberg game is approximated with a sequence of sub-minimax problem, and we prove that the sub-minimax problem can be efficiently solved by the convex optimization even though it belongs to the convex non-concave optimization problems. Finally, numerical results highlight the
effectiveness and competitiveness of the proposed algorithms as well as the optimized waveform-filter pair.

\end{abstract}


%
\IEEEpeerreviewmaketitle

\section{Introduction}
%
%
%
%

\IEEEPARstart{T}{he} past decades have witnessed the great development of Multiple-Input Multiple-Output(MIMO) radar which improves the parameter estimation and target detection performance significantly by utilizing waveform diversity{\cite{LiMIMO,LiMIMO2008,LiOnparameter,bliss2003multiple}. Owing to the fact that targets usually exhibits range-spread characteristics for high resolution radar, waveform design for extended targets has drawn increasing attention from researchers during the past decades{\cite{karbasi2015robust,bell1993information,yang2007mimo,meng2012radar,tang2016robust,yao2020robust}}.

Given that the performance of target detection and parameter estimation depends on the output Signal-to-Interference-plus-Noise Ratio(SINR), a preferred waveform optimization criterion is maximizing output SINR at the receiver\cite{bell1993information,chen2009mimo,jiu2012minimax,tang2016robust,aubry2013knowledge,cheng2017mimo,stoica2011optimization,pillai1999optimum}. In \cite{bell1993information}, Bell studies the matched illumination waveform for the extended target. As stated in \cite{pillai1999optimum}, the authors investigate joint design of transmit waveform and receive filter via maximizing the output SINR. However, the aforementioned waveform design algorithms need the exact knowledge of Target Impulse Response(TIR) or Power Spectral Density(PSD), which is often used to characterize scattering behavior of extended target{\cite{kay2007optimal,li1996scattering}. Unfortunately, TIR is sensitive to a variety of factors, and it is impossible to get the exact knowledge of TIR when designing radar waveform. To this end, robust waveform design is taken into account leading to the minimax optimization problem\cite{razaviyayn2020nonconvex,chen2009mimo,jiu2012minimax,yao2020robust}. In most cases, the uncertainty of TIR is modeled as a scaled sphere centered around a known but imprecise TIR(i.e., see \cite{chen2009mimo,karbasi2015robust,tang2016robust,yao2020robust} and references therein). Aiming at maximizing the worst-case SINR under the spherical uncertainty of TIR, \cite{chen2009mimo} proposes an iterative algorithm to optimize transmit waveform and receive filter alternatively. And \cite{tang2016robust} addresses the problem based on the minimax theorems\cite{sion1958general}. Nevertheless, the algorithms proposed in \cite{chen2009mimo} and \cite{tang2016robust} can only deal with the energy constrained waveform, which might result in the undesired waveform for radar with a high Peak-to-Average power Ratio(PAR). In order to avoid this drawback, \cite{karbasi2015robust} imposes a PAR constraint on the waveform and proposes the alternative optimization algorithm with respect to the covariance of waveform and filter, where the spherical uncertainty of TIR is approximated by random samples from its surface. To our knowledge, there is few literature considering other constraints on transmit waveform, such as similarity or spectral compatibility constraint\cite{cui2017quadratic,cui2013mimo,aubry2016optimization}, for the worst-case situation.   
	
Game theory, modeling the interaction of players, is applied in a broad variety of fields, such as economics, social science and machine learning\cite{morgenstern1953theory,myerson2013game,goodfellow2014generative}. Recently, game theory methods have been introduced to the context of signal processing for jamming suppression and waveform design\cite{han2016jointly,panoui2016game,zhang2019game,song2011mimo}. As to the extended target detection problem, it can be modeled as a Two-Person Zero-Sum game(TPZS) if the target is "smart" enough and always tries to prevent being detected by radar. Apart from this, the interaction between radar and target forms a Stackelberg game if one acts as the leader, for example, the robust waveform design problem mentioned in \cite{chen2009mimo,karbasi2015robust} can be regarded as the Stackelberg game where radar acts as the leader. And the robust waveform can be considered as the equilibrium strategy of radar in the Stackelberg game.

In this paper, we consider the MIMO radar waveform-filter design problem by constructing Stackelberg game between radar and the extended target with SINR being the payoff function. The goal is to obtain the Stackelberg equilibrium strategy of radar under several practical constraints on waveform, namely, Energy Constraint(EC), Constant Modulus and Similarity Constraint(CM-SC), Spectral Compatibility and Similarity Constraint(SC-SC). Furthermore, under some special conditions, we will prove that the Stackelberg equilibrium is also the Nash equilibrium for the game and the optimal waveform-filter pair is calculated by solving Nash equilibrium. Generally speaking, our work makes the following contributions:

1) \textit{The Nash equilibrium under EC on waveform}: For the case of EC on waveform, we theoretically prove that the Stackelberg equilibrium is also the Nash equilibrium, and propose \textbf{Algorithm 1} to solve it through convex optimization. Even though, the relevant researches are reported in \cite{tang2016robust}, the proof procedure is imperfect since the Stackelberg game in this case belongs to the non-convex concave minimax problem and can't be converted to the convex concave form by diagonal loading. Consequently, we prove it from a totally new view by utilizing the optimality condition, which has  been seldom reported in open literature.

2) \textit{The Nash equilibrium for the (Semi-Definite Programming)SDP relaxation form under CM-SC on waveform}: In the case of CM-SC on waveform, the Nash equilibrium might not exist for the TPZS game between radar and target. But we prove that the Nash equilibrium always exists for its SDP relaxation based on Sion's minimax theorem\cite{sion1958general}. And we devise \textbf{Algorithm 2} to calculate the Nash equilibrium approximately. Finally, the optimal waveform-filter is synthesized from the strategy of Nash equilibrium with randomization.

3) \textit{The Stackelberg equilibrium under SC-SC on waveform}: 
\textbf{Algorithm 3} is proposed to solve the Stackelberg equilibrium under SC-SC on waveform leveraging on (Majorization Minimization)MM algorithm. Based on the dual theorem\cite{BoydConvex}, we prove that the minimax problem at each iteration of MM algorithm can be solved through convex optimization, even though it is convex non-concave. Thus, the Stackelberg equilibrium can be calculated in polynomial time.  

4) \textit{Analyses and experiments for proposed algorithms}: The convergences as well as computational complexities of the proposed 3 algorithms are analyzed, respectively. Numerical experiments are carried out to verify the effectiveness of the proposed algorithms. In addition, some comparisons with the current algorithms are given. The results highlight the superiority of our algorithms to some extend. 

The remainder of this paper is organized as follows. Section II establishes the MIMO radar signal model and formulates the TPZS game between the radar and extended target. Section III is devoted to the algorithms and analyses for equilibrium strategies under different waveform constraints. Section IV provides several numerical experiments and analyses of the proposed algorithms, and exhibits the performance of the waveform-filter pair. Finally, Section V concludes the paper.

\textit{Notations:} Throughout this paper, scalars are denoted by italic letters(e.g., \textit{a}, \textit{A}); vectors are denoted by bold italic lowercase letters(e.g., $\bm{a}$) and $\bm{a}(i)$ denotes the $i$th element of $\bm{a}$; matrices are denoted by bold italic capital letters(e.g., $\bm{A}$) and  $\bm{A}(i,j)$ denotes the element in the $i$th row and $j$th column of $\bm{A}$. Superscript ${\bar{\left( \cdot\right)}}$, ${\left( \cdot\right) ^{\rm{T}}}$ and ${\left( \cdot\right) ^{\rm{H}}}$ denote complex conjugate, transpose and conjugate transpose, respectively. tr($\cdot$) denotes the trace of a square matrix. vec($\cdot$) denotes the operator of column-wise stacking a matrix. $\otimes$ and $\odot$ represent Kronecker product and Hadamard product, respectively. 
diag($\bm{a}$) denotes the diagonal matrix with the diagonal elements formed by $\bm{a}$, while diag($\bm{A}$) denotes the vector with elements formed by the diagonal elements of $\bm{A}$. $\left\langle \cdot, \cdot \right\rangle$, $\left\| \cdot \right\| _2$, $\left\| \cdot \right\| _\infty$  and $\left\| \cdot \right\| _F$ denote the inner product, $l_2$ norm, $l_\infty$ norm and Frobenius norm in Euclidean space, respectively. arg(${a}$) and $\left| {a}\right|$ denote the phase angle and the modulus of ${a}$, respectively. The representation $\bm{A} \succ 0(\bm{A}\succeq 0)$ means $\bm{A}$ is positive definite(semi-definite).      

\section{Signal Model and Problem Formulation }

\subsection{Signal model}
As outlined in Fig.\ref{Scenario}, the mission of radar is to detect an extended target at direction ${\theta _{t}}$, and the extended target tries to prevent being detected by adjusting its attitude within a certain range, which affects its returns to radar. Then, the returns received by radar are given by 
\begin{equation}
{\bm{y}} = {{\bm{y}}_t} +  {\bm{n}},
\end{equation}
if the clutter is neglected, where ${{\bm{y}}_t}$ and ${\bm{n}}$ denote the extended target returns and noise, respectively.

Now, we are going to derive the signal model of ${{\bm{y}}_t}$. Let us consider a colocated MIMO radar with ${N_T}$ transmitters and ${N_R}$ receivers, assuming that the waveform transmitted by the \textit{n}th transmitter with \textit{L} samples is denoted by ${{\bm{s}}_n} = {[{{\bm{s}}_n}(1),{{\bm{s}}_n}(2),...,{\bm{s}}_n(L)]^{\rm{T}}}$, then the transmitting matrix for MIMO array can be represented as ${\bm{S}} = {[{{\bm{s}}_1},{{\bm{s}}_2},...,{{\bm{s}}_{{N_T}}}]^{\rm{T}}} \in {\mathbb{C}}^{{N_T} \times L}$. In the far field, the signal incident on the target can be written as ${{\bm{a}}^{\rm{T}}}({\theta _t}){\bm{S}}$, where ${{\bm{a}}({\theta _t})}$ is the transmit array steering vector at ${\theta _t}$.

Denote by $\bm{t} \in {\mathbb{C}}^{Q}$ the TIR of target, and the returns modulated by it can be represented as  
\begin{equation}
{\bm{T}}{\left( {{{\bm{a}}^{\rm{T}}}({\theta _t}){\bm{S}}} \right)^{\rm{T}}} = {\bm{T}}{{\bm{S}}^{\rm{T}}}{\bm{a}}({\theta _t}),
\end{equation}
where ${\bm{T}} = \sum\limits_{i = 1}^Q {{\bm{t}}(i){{\bm{J}}_{i - 1}}}$ is the TIR matrix with ${{\bm{J}}_{i}}$ being the $(Q+L-1) \times L$ shift matrix given by\cite{karbasi2015robust} 
\begin{equation}
{{\bm{J}}_i}(m,n) = \left\{ \begin{array}{l}
1\;,\quad m - n = i\\
0,\;\quad m - n \ne i
\end{array} \right..
\end{equation}

Consequently, one get the signal matrix $\bm{Y}_t$ at the MIMO receiver with
\begin{equation}
{{\bm{Y}}_t} = {\bm{b}}({\theta _t}){{\bm{a}}^{\rm{T}}}({\theta _t}){\bm{S}}{{\bm{T}}^{\rm{T}}},
\label{SingalModelMatrixForm}
\end{equation}
where ${\bm{b}}({\theta _t})$ denotes the receive steering vector at ${\theta _t}$. Let ${\bm{s}} = {\rm{vec}}({\bm{S}})$, the vectorization form of $\bm{Y}_t$ is given by
\begin{equation}
{{\bm{y}}_t} = \left( {{\bm{T}} \otimes \left( {{\bm{b}}({\theta _t}){{\bm{a}}^{\rm{T}}}({\theta _t})} \right)} \right){\bm{s}} = \bm{G}(\bm{t})\bm{s}.
\label{SingalModelVectorFormS}
\end{equation}
After a couple of matrix operations, it is easy to get another expression form of (\ref{SingalModelVectorFormS}) as follow,
\begin{equation}
{{\bm{y}}_t} = {\bm{H}}({\bm{s}}){\bm{t}},
\label{SingalModelVectorFormT}
\end{equation}
where ${\bm{H}}({\bm{s}}) = \left[ {{{\bm{h}}_1}({\bm{s}}),{{\bm{h}}_2}({\bm{s}}),...,{{\bm{h}}_Q}({\bm{s}})} \right]$ with \vspace{2pt} ${{\bm{h}}_i}({\bm{s}}) = \left( {{{\bm{J}}_{i-1}} \otimes \left( {{\bm{b}}({\theta _t}){{\bm{a}}^{\rm{T}}}({\theta _t})} \right)} \right){\bm{s}}={\bm{A}_i}\bm{s}$.

As to the noise, it is assumed to be complex Gaussian with ${\bm{n}}\sim\mathcal{CN}({\bf{0}},{{\bm{R}}_c})$.

%
%
 
\begin{figure}[!t]
  	\centering
 	\includegraphics[width=3.5in,height=2.6in]{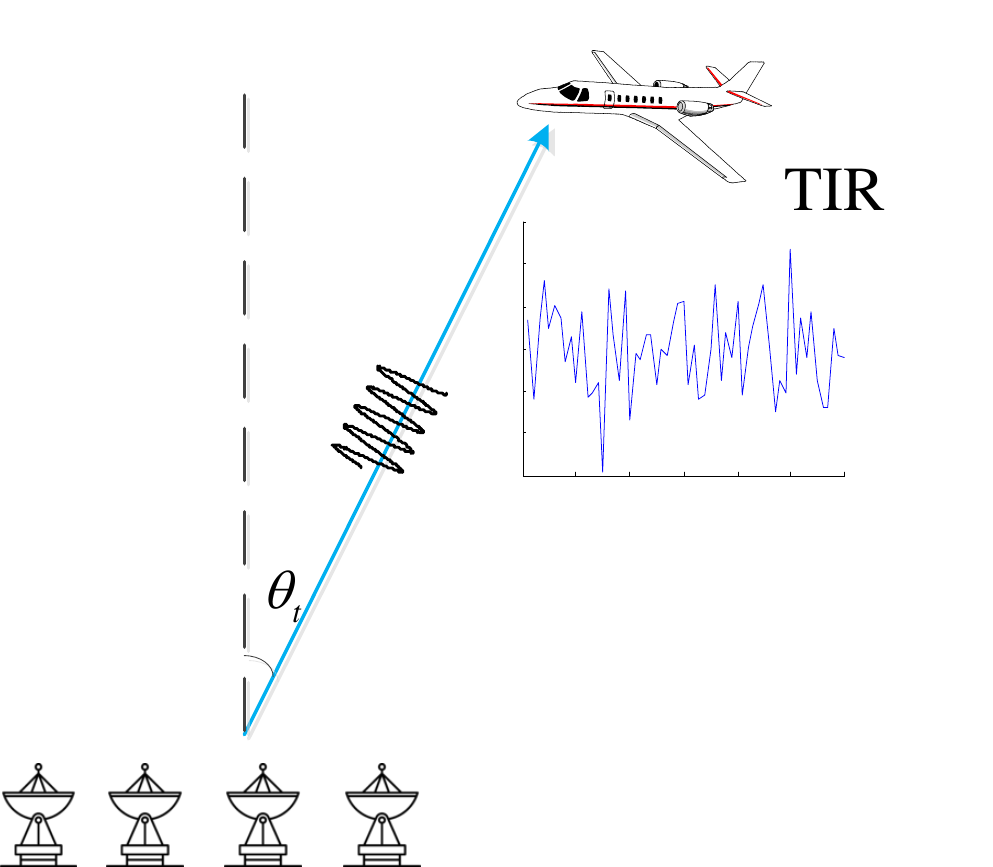}
  \caption{Extended target detection diagram}
  \label{Scenario}
\end{figure}
\subsection{Games between radar and target}
The radar processor extract the desired information at the receiver with a linear filter $\bm{w}$ achieving the output SINR as
\begin{equation}
{\rm{SINR}} = \frac{{{{\left| {{{\bm{w}}^{\rm{H}}}{{\bm{y}}_t}} \right|}^2}}}{{{{\bm{w}}^{\rm{H}}}{{\bm{R}}_c}{\bm{w}}}}.
\end{equation}

For a given probability of false alarm $P_{fa}$, analytical form of the detection probability $P_{d}$ is given by \cite{richards2005fundamentals,de2011design,kay1993fundamentals}
\begin{equation}
{P_d}{\rm{ = }}Q\left( {\sqrt {2{\rm{SINR}}} {\rm{,}}\sqrt { - 2\log ({P_{fa}})} } \right),
\label{MarcumQFunction}
\end{equation}
where $Q\left( {\cdot {\rm{,}}\cdot}  \right)$ denotes the Marcum-Q function.
 
Consider the TPZS game \cite{morgenstern1953theory,myerson2013game} between the radar and the target with a common payoff function $\rm{SINR}$ with respect to ($\bm{s}$,$\bm{w}$,$\bm{t}$). The radar player tries to maximize the SINR to capture the target with high probabilities by choosing the waveform-filter pair ($\bm{s}$,$\bm{w}$) from its strategy set $\Psi$. Alternatively, the target player tries to minimize the SINR to avoid being detected by choosing the TIR from its strategy set $\Omega$. 

Assume that the radar player goes first and the target player is able to perceive the radar's strategy. Then, the interaction between radar and target results in the following Stackelberg game where the radar player acts as the leader,  
\begin{equation}
{{\cal P}_r}\left\{ \begin{array}{l}
\underline {{\rm{SINR}}}  = \mathop {{\rm{max}}}\limits_{{\bm{s}},{\bm{w}}} \mathop {{\rm{ min}}}\limits_{\bm{t}} {\rm{ }}\frac{{{{\left| {{{\bm{w}}^{\rm{H}}}{{\bm{y}}_t}} \right|}^2}}}{{{{\bm{w}}^{\rm{H}}}{{\bm{R}}_c}{\bm{w}}}}\\
s.t.\quad \left( {{\bm{s}},{\bm{w}}} \right) \in \Psi ,{\bm{t}} \in \Omega 
\end{array} \right..  
\label{RadarOptimalStratagy}
\end{equation}

Conversely, assume that the target player goes first and the radar player is able to perceive the target's strategy. Then, the interaction between radar and target results in the following Stackelberg game where the target player acts as the leader,
\begin{equation}
{{\cal P}_t}\left\{ \begin{array}{l}
\overline {{\rm{SINR}}}  = \mathop {{\rm{min}}}\limits_{\bm{t}} \mathop {{\rm{max}}}\limits_{{\bm{s}},{\bm{w}}} {\rm{ }}\frac{{{{\left| {{{\bm{w}}^{\rm{H}}}{{\bm{y}}_t}} \right|}^2}}}{{{{\bm{w}}^{\rm{H}}}{{\bm{R}}_c}{\bm{w}}}}\\
s.t.\quad \left( {{\bm{s}},{\bm{w}}} \right) \in \Psi ,{\bm{t}} \in \Omega 
\end{array} \right..
\label{TargetOptimalStratagy}
\end{equation}

The following weak minimax inequality
\begin{equation}
\underline {{\rm{SINR}}}  \le \overline {{\rm{SINR}}}
\label{WeakMinimaxInequality}
\end{equation}
always holds, which can be regarded as the information gain for the player who plays second.  However, in some certain conditions, the strong minimax inequality\cite{sion1958general}
\begin{equation}
\underline {{\rm{SINR}}}  = \overline {{\rm{SINR}}}
\label{StrongMinimaxInequality}
\end{equation}
holds, which means that it makes no difference for the player who goes first or second. And the quantity ${{\rm{SINR}}}^* =  \underline {{\rm{SINR}}}= \overline {{\rm{SINR}}}$ is called the value of the TPZS game.

It is worth pointing out that the problem ${{\cal P}_r}$ is also referred as robust or minimax waveform-filter design in some literatures\cite{chen2009mimo,jiu2012minimax,karbasi2015robust}.
Moreover, denote by $(\bm{s}^*,\bm{w}^*)$ the Stackelberg equilibrium strategy of radar for ${{\cal P}_r}$, and $\bm{t}^*$ the Stackelberg equilibrium strategy of target for ${{\cal P}_t}$, respectively. Then, the strategy pair $(\bm{s}^*,\bm{w}^*,\bm{t}^*)$ is also a Nash equilibrium \cite{myerson2013game} for the TPZS game, if the equality (\ref{StrongMinimaxInequality}) holds. To this end, in the following paper, we focus on solving the optimal solution $(\bm{s}^*,\bm{w}^*)$, namely Stackelberg or Nash equilibrium strategies of radar in the TPZS game in different cases.

\section{Equilibrium Strategies and Optimal Waveform-Filter Pair}
This section is devoted to the equilibrium strategies as well as the optimal waveform-filter pair of radar for the TPZS game under different constraints on waveform. In particular, we assume that the target strategies are bounded in a scaled sphere centered around a previous known $\bm{t}_0$, namely,
\begin{equation}
\Omega  = \left\{ {{\bm{t}}|{{\left\| {{\bm{t}} - {{\bm{t}}_0}} \right\|}_2} \le r} \right\}.
\end{equation}
As to the radar player, we assume that the filter $\bm{w}$ is unconstrained and the waveform belongs to the following three sets:
\noindent $\bullet$ EC set 
\begin{equation}
{\Psi _E} = \left\{ {{{\bm{s}}} |{{\left\| {\bm{s}} \right\|}_2^2} \le e_t,} \right\},
\end{equation}
where $e_t$ is the total available power;

\noindent $\bullet$ CM-SC set
\begin{equation}
{\Psi _C} = \left\{ {{\bm{s}}|\left| {{\bm{s}}(i)} \right| = \sqrt{\frac{{ {{e_t}} }}{{{{N_T}L} }}},{{\left\| {{\bm{s}} - {{\bm{s}}_0}} \right\|}_\infty } \le \frac{{\delta \sqrt {{e_t}} }}{{\sqrt {{N_T}L} }}} \right\},
\end{equation}
where $\bm{s}_0$ is a reference signal with good ambiguity properties, and $\delta \leq 2$ controls the similarly between $\bm{s}$ and $\bm{s}_0$;

\noindent $\bullet$ SC-SC set
\begin{equation}
{\Psi _S} = \left\{ { {{\bm{s}}} |{{\left\| {\bm{s}} \right\|}_2^2} \leq  e_t,{{\left\| {{\bm{s}} - {{\bm{s}}_0}} \right\|}_\infty } \le \frac{{\delta \sqrt {{e_t}} }}{{\sqrt {{N_T}L} }},{{\bm{s}}^{\rm{H}}}{{\bm{R}}_I}{\bm{s}} \le {e_I}} \right\},
\end{equation}
where $e_I$ denotes the maximum allowed energy allocated on the specified frequency bands, and $\bm{R}_I$ is the corresponding spectrum compatibility matrix defined as ${{\bm{R}}_I} = \sum\limits_{k = 1}^K {{\alpha _k}{{\bf{I}}_{{N_T}}} \otimes {\bm{R}}_I^k}$ with
\begin{equation}
{\bm{R}}_I^k(m,n) = \left\{ \begin{array}{l}
f_2^k - f_1^k,\;\;m = n\\
\frac{{{e^{j2\pi f_2^k(m - n)}} - {e^{j2\pi f_1^k(m - n)}}}}{{j2\pi (m - n)}},m \ne n
\end{array} \right.,
\end{equation}
where $[f_1^k,f_2^k]$ denotes the $k$th normalized frequency interval of transmission and ${\alpha _k} \ge 0$ is the weight for the $k$th frequency band\cite{aubry2015new,tang2016joint}.

Under this assumption, the waveform-filter pair design problem belongs to a class of non-convex concave minimax problems. Nevertheless, as will be illustrated later, some of these problems can be solved or approximately solved by constructing the Nash equilibrium between the two players. To this end, let's provide the basic theorem in our paper.
\newtheorem{theorems}{Theorem}
\begin{theorems}
Let $\Omega$ be a convex set and $\bm{w}$ be unconstrained, then ${\cal{P}}_r$ is equivalent\footnote{By "equivalent", we mean that the optimal solutions can be constructed from each other} to 
\begin{equation}
{\tilde {\cal P}_r}\left\{ \begin{array}{l}
 \mathop {{\rm{max}}}\limits_{\bm{s}} \mathop {{\rm{min}}}\limits_{\bm{t}} \mathop {{\rm{max}}}\limits_{\bm{w}} {\rm{ }}\frac{{{{\left| {{{\bm{w}}^{\rm{H}}}{{\bm{y}}_t}} \right|}^2}}}{{{{\bm{w}}^{\rm{H}}}{{\bm{R}}_c}{\bm{w}}}}\\
s.t.\quad {\bm{s}} \in \Psi ,{\bm{t}} \in \Omega 
\end{array} \right.
\end{equation}
\end{theorems}
\begin{IEEEproof}
	See Appendix A.
\end{IEEEproof}

As an immediate consequence of \textit{Theorem 1}, we can reformulate ${\cal{P}}_r$ and ${\cal{P}}_t$ by solving the inner maximization problem with $\bm{w}=\bm{R}_c^{-1}\bm{y}_t$ as 
\begin{equation}
{{\cal P}_r}\left\{ \begin{array}{l}
\underline {{\rm{SINR}}}  = \mathop {{\rm{max}}}\limits_{\bm{s}} \mathop {{\rm{ min}}}\limits_{\bm{t}} {\rm{ }}{\bm{y}}_t^{\rm{H}}{\bm{R}}_c^{ - 1}{{\bm{y}}_t}\\
s.t.\quad  {\bm{s}} \in \Psi ,{\bm{t}} \in \Omega 
\end{array} \right.
\end{equation}
and 
\begin{equation}
{{\cal P}_t}\left\{ \begin{array}{l}
\overline {{\rm{SINR}}}  = \mathop {{\rm{min}}}\limits_{\bm{t}} \mathop {{\rm{max}}}\limits_{\bm{s}} {\rm{ }}{\bm{y}}_t^{\rm{H}}{\bm{R}}_c^{ - 1}{{\bm{y}}_t}\\
s.t.\quad {\bm{s}} \in \Psi ,{\bm{t}} \in \Omega 
\end{array} \right..
\end{equation}

\subsection{EC on waveform}
In this subsection, we discuss the optimal solutions of ${\cal{P}}_r$ and ${\cal{P}}_t$ with the case $\Psi = \Psi_E$. Substituting $\Psi_E$ into ${\cal{P}}_r$ and ${\cal{P}}_t$, we recast the game as 
\begin{equation}
{{\cal P}_r^E}\left\{ \begin{array}{l}
\underline {{\rm{SINR}}}_E  = \mathop {{\rm{max}}}\limits_{\bm{s}} \mathop {{\rm{ min}}}\limits_{\bm{t}} {\rm{ }}{\bm{y}}_t^{\rm{H}}{\bm{R}}_c^{ - 1}{{\bm{y}}_t}\\
s.t.\quad \left\| {\bm{s}} \right\|_2^2 \le {e_t},{\left\| {{\bm{t}} - {{\bm{t}}_0}} \right\|_2} \le r
\end{array} \right.
\end{equation}
and 
\begin{equation}
{{\cal P}_t^E}\left\{ \begin{array}{l}
\overline {{\rm{SINR}}}_E  = \mathop {{\rm{min}}}\limits_{\bm{t}} \mathop {{\rm{max}}}\limits_{\bm{s}} {\rm{ }}{\bm{y}}_t^{\rm{H}}{\bm{R}}_c^{ - 1}{{\bm{y}}_t}\\
s.t.\quad \left\| {\bm{s}} \right\|_2^2 \le {e_t},{\left\| {{\bm{t}} - {{\bm{t}}_0}} \right\|_2} \le r
\end{array} \right..
\end{equation}
And the following proposition demonstrates the existence of Nash equilibrium for the game.
\newtheorem{propositions}{Proposition}
\begin{propositions}
Let $\bm{s}_E^*$ and $\bm{t}_E^*$ are the optimal solutions of  ${{\cal{P}}_r^E}$ and ${{\cal{P}}_t^E}$, respectively. Then, the strategy pair $(\bm{s}_E^*,\bm{t}_E^*)$ is Nash equilibrium for the TPZS game and  $\underline {{\rm{SINR}}}_E = \overline {{\rm{SINR}}}_E$. 
\end{propositions}
\begin{IEEEproof}
	See Appendix B.
\end{IEEEproof}
 
According to \textit{Proposition 1}, one can get the idea that the optimal waveform can be obtained by solving either ${\cal{P}}_r^E$ or ${\cal{P}}_t^E$. Obviously, solving ${\cal{P}}_t^E$ is a much better choice since it can be converted to a convex optimization problem. Note that  ${\cal{P}}_t^E$ is equivalent to the following problem by substituting (\ref{SingalModelVectorFormS}) into it
\begin{equation}
{\cal P}_t^E\left\{ \begin{array}{l}
\mathop {{\rm{min}}}\limits_{\bm{t}} \mathop {{\rm{max}}}\limits_{\bm{s}} {\rm{ }}{{\bm{s}}^{\rm{H}}}{\bm{G}}{({\bm{t}})^{\rm{H}}}{\bm{R}}_c^{ - 1}{\bm{G}}({\bm{t}}){\bm{s}}\\
s.t.\quad \left\| {\bm{s}} \right\|_2^2 \le {e_t},{\left\| {{\bm{t}} - {{\bm{t}}_0}} \right\|_2} \le r
\end{array} \right.,
\end{equation}
and the analytical form of the optimal solution with respect to $\bm{s}$ is 
\begin{equation}
{\bm{s}_E^*} = \sqrt{e_t}{\cal{M}}\left( {{\bm{G}}{{({\bm{t}})}^{\rm{H}}}{\bm{R}}_c^{ - 1}{\bm{G}}({\bm{t}})} \right),
\label{OptimalWaveformEigenVector}
\end{equation}
where ${\cal{M}}(\bm{A})$ denotes the normalized principle eigenvector of $\bm{A}$. Then, it reduces to 
\begin{equation}
\tilde {\cal P}_t^E\left\{ \begin{array}{l}
\mathop {{\rm{min}}}\limits_{\bm{t}} {\rm{ }}{\lambda _{\max }}\left( {{\bm{G}}{{({\bm{t}})}^{\rm{H}}}{\bm{R}}_c^{ - 1}{\bm{G}}({\bm{t}})} \right)e_t\\
s.t.\quad {\left\| {{\bm{t}} - {{\bm{t}}_0}} \right\|_2} \le r
\end{array} \right.,
\label{OptimizationLambda}
\end{equation} 
where $\lambda_{\max }(\bm{A})$ denotes the maximum eigenvalue of $\bm{A}$. 

Based on Schur complement theorem\cite{BoydConvex}, we recast  $\tilde {\cal P}_t^E$ as follow 
\begin{equation}
\hat {\cal P}_t^E\left\{ \begin{array}{l}
\mathop {{\rm{min}}}\limits_{\bm{t},\mu} {\rm{ }}e_t\mu \\
s.t.\quad \left[ {\begin{array}{*{20}{c}}
	{\mu {\bf{I}}}&{{\bm{G}}{{({\bm{t}})}^{\rm{H}}}}\\
	{{\bm{G}}({\bm{t}})}&{{{\bm{R}}_c}}
	\end{array}} \right] \succeq 0,{\left\| {{\bm{t}} - {{\bm{t}}_0}} \right\|_2} \le r
\end{array} \right.,
\end{equation}
where $\mu$ is the auxiliary variable.

It is worth pointing out that $\bm{G}(\bm{t})$ is a affine function with respect to $\bm{t}$ and the SDP problem can be solved efficiently in polynomial time by CVX\cite{GrantCVX}. Once the optimal solution $\bm{t}_E^*$ is obtained, we can obtain the optimal waveform-filter pair  $(\bm{s}_E^*,\bm{w}_E^*)$  with
\begin{equation}
{\bm{s}}_E^* = \sqrt{e_t}{\cal{M}}\left( {{\bm{G}}{{({\bm{t}}_E^*)}^{\rm{H}}}{\bm{R}}_c^{ - 1}{\bm{G}}({\bm{t}}_E^*)} \right),{\bm{w}}_E^* = {\bm{R}}_c^{ - 1}{\bm{H}}({\bm{s}}_E^*){\bm{t}}_E^*.
\label{OptimalWaveformFilterEC}
\end{equation}  

The overall procedure for waveform-filter design under EC is summarized in Table \ref{Algorithm1EC} as \textbf{Algorithm 1}. The convergence of \textbf{Algorithm 1} is guaranteed by the convexity of ${\cal{\hat{P}}}_t^E$. As to the computational complexities, 
it requires at most $O\left( {{Q{\left( {L{N_T} + L{N_R} + Q{N_R}} \right)}^{3.5}}} \right)$ to solve ${\cal{\hat{P}}}_t^E$\cite{NemirovskiLectures} and $O\left( {{{\left( {L{N_T}} \right)}^3}} \right) + O\left( {\left( {L{N_R} + Q{N_R}} \right)^2} \right)$ operations to calculate ($\bm{s}_E^*,\bm{w}_E^*$) with  (\ref{OptimalWaveformFilterEC}). Therefore, the total \vspace{2pt}computational complexities will not exceed $O\left( {{Q{\left( {L{N_T} + L{N_R} + Q{N_R}} \right)}^{3.5}}} \right)$.

\begin{table}[!t]
	\renewcommand{\arraystretch}{1.3}
	\caption{Algorithm for waveform-filter optimization with EC}
	\label{Algorithm1EC}
	\centering
	\begin{tabular}{l}
		\hline
		\textbf{Input}:
		$\bm{a}(\theta_{t})$, $\bm{b}(\theta_{t})$, $e_t$,  
		${{\bm{R}}_{c}}$, ${{\bm{t}}_{0}}$ and $r$.\\ 
		\hline
		\@ \@\@ \@ \textit{Step 1}: Solve ${\cal{\hat{P}}}_t^E$ and obtain $\bm{t}_E^*$.\\
		\@ \@\@ \@ \textit{Step 2}: Synthesize ${\bm{s}_E^{*}}$ and ${\bm{w}_E^{*}}$ with (\ref{OptimalWaveformFilterEC}).\\
		\hline
		\textbf{Output}: 
		\@ \@ ${\bm{s}_E^{*}}$ and ${\bm{w}_E^{*}}$.\\
		\hline
	\end{tabular}
\end{table}

\subsection{CM-SC on waveform} 
In this subsection, we study ${\cal{P}}_r$ and ${\cal{P}}_t$ with the case that $\Psi = \Psi_C$. Now, we recast the game as

\begin{equation}
{\cal{P}}_r^C\left\{ {\begin{array}{*{20}{l}}
\underline {{\rm{SINR}}}_C = {\mathop {{\rm{max}}}\limits_{\bm{s}} \mathop {{\rm{min}}}\limits_{\bm{t}} {{\bm{s}}^{\rm{H}}}{\bm{G}}{{({\bm{t}})}^{\rm{H}}}{\bm{R}}_c^{ - 1}{\bm{G}}({\bm{t}}){\bm{s}}}\\
{s.t.\;\;\;{\kern 1pt} \left| {\bm{s}}(i)\right|  = \sqrt {\frac{{{e_t}}}{{{N_T}L}}} ,{{\left\| {{\bm{s}} - {{\bm{s}}_0}} \right\|}_\infty } \le \frac{{\delta \sqrt {{e_t}} }}{{\sqrt {{N_T}L} }}}\\
{\;\;\;{\kern 1pt} \;\;\;{\kern 1pt} {{\left\| {{\bm{t}} - {{\bm{t}}_0}} \right\|}_2} \le r}
\end{array}} \right.
\end{equation}
and

\begin{equation}
{\cal{P}}_t^C\left\{ {\begin{array}{*{20}{l}}
\overline {{\rm{SINR}}}_C={\mathop {{\rm{min}}}\limits_{\bm{t}} \mathop {{\rm{max}}}\limits_{\bm{s}} {{\bm{s}}^{\rm{H}}}{\bm{G}}{{({\bm{t}})}^{\rm{H}}}{\bm{R}}_c^{ - 1}{\bm{G}}({\bm{t}}){\bm{s}}}\\
{s.t.\;\;\;{\kern 1pt} \left|  {\bm{s}}(i) \right| = \sqrt {\frac{{{e_t}}}{{{N_T}L}}} ,{{\left\| {{\bm{s}} - {{\bm{s}}_0}} \right\|}_\infty } \le \frac{{\delta \sqrt {{e_t}} }}{{\sqrt {{N_T}L} }}}\\
{\;\;\;{\kern 1pt} \;\;\;{\kern 1pt} {{\left\| {{\bm{t}} - {{\bm{t}}_0}} \right\|}_2} \le r}
	\end{array}} \right..
\end{equation}

Note that the objective function in ${\cal{P}}_r^C$ and ${\cal{P}}_r^C$ is convex with respectively to $\bm{t}$, but non-concave with respect to $\bm{s}$, thus, it is always impossible to find a satisfactory solution in
polynomial time. To this end, we study its relaxation form by letting $\bm{R}_s = \bm{s}\bm{s}^{\rm{H}}$ and dropping the rank as well as the similarity constraint, which leads to the following SDP problems
\begin{equation}
 \tilde{{\cal{P}}}_r^C\left\{ \begin{array}{l}
\mathop {{\rm{max}}}\limits_{{{\bm{R}}_s}} \mathop {{\rm{min}}}\limits_{\bm{t}} {\rm{trace}}\left( {{\bm{G}}{{({\bm{t}})}^{\rm{H}}}{\bm{R}}_c^{ - 1}{\bm{G}}({\bm{t}}){{\bm{R}}_s}} \right)\\
s.t.\;\;\;{\kern 1pt} {{\bm{R}}_s}(i,i) = \frac{{{e_t}}}{{{N_T}L}},\bm{R}_s \succeq 0,\\
\;\quad \quad {\left\| {{\bm{t}} - {{\bm{t}}_0}} \right\|_2} \le r
\end{array} \right.
\label{RadarPlayerSDP}
\end{equation}
and
\begin{equation}
 \tilde{{\cal{P}}}_t^C\left\{ \begin{array}{l}
\mathop {{\rm{min}}}\limits_{{{\bm{t}}}} \mathop {{\rm{max}}}\limits_{\bm{R}_s} {\rm{trace}}\left( {{\bm{G}}{{({\bm{t}})}^{\rm{H}}}{\bm{R}}_c^{ - 1}{\bm{G}}({\bm{t}}){{\bm{R}}_s}} \right)\\
s.t.\;\;\;{\kern 1pt} {{\bm{R}}_s}(i,i) = \frac{{{e_t}}}{{{N_T}L}},\bm{R}_s \succeq 0,\\
\;\quad \quad {\left\| {{\bm{t}} - {{\bm{t}}_0}} \right\|_2} \le r
\end{array} \right..
\label{TargetPlayerSDP}
\end{equation}

Now, we turn to the properties of $\tilde{{\cal{P}}}_r^C$ and $\tilde{{\cal{P}}}_t^C$. For a fixed $\bm{t}$, it is easy to verify that ${z^C}\left( {{{\bm{R}}_s},{\bm{t}}} \right) = {\rm{trace}}\left( {{\bm{G}}{{({\bm{t}})}^{\rm{H}}}{\bm{R}}_c^{ - 1}{\bm{G}}({\bm{t}}){{\bm{R}}_s}} \right)$ is linear with respect to $\bm{R}_s$. On the other hand, based on (\ref{SingalModelVectorFormT}) we can derive another expression of ${z^C}\left( {{{\bm{R}}_s},{\bm{t}}} \right)$, given by
\begin{equation}
{z^C}\left( {{{\bm{R}}_s},{\bm{t}}} \right) = {{\bm{t}}^{\rm{H}}}{\bm{H}}{({\bm{s}})^{\rm{H}}}{\bm{R}}_c^{ - 1}{\bm{H}}({\bm{s}}){\bm{t}} = {{\bm{t}}^{\rm{H}}}{\bm{U}}({{\bm{R}}_s}){\bm{t}}
\end{equation}
with 
\begin{equation}
{\bm{U}}({{\bm{R}}_s})(i,j) = {{\bm{s}}^{\rm{H}}}{\bm{A}}_i^{\rm{H}}{\bm{R}}_c^{ - 1}{\bm{A}}_i^{}{\bm{s}} = {\rm{trace}}\left( {{\bm{A}}_i^{\rm{H}}{\bm{R}}_c^{ - 1}{\bm{A}}_i^{}{{\bm{R}}_s}} \right).
\end{equation}
As a result, for a fixed $\bm{R}_s$,  ${z^C}\left( {{{\bm{R}}_s},{\bm{t}}} \right)$ is a convex quadratic form with respect to $\bm{t}$. According to Sion's theorem in \cite{sion1958general}, we have 
\begin{equation}
\begin{array}{l}
\mathop {{\rm{min}}}\limits_{\bm{t}} \mathop {{\rm{max}}}\limits_{{{\bm{R}}_s}} {\rm{ trace}}\left( {{\bm{G}}{{({\bm{t}})}^{\rm{H}}}{\bm{R}}_c^{ - 1}{\bm{G}}({\bm{t}}){{\bm{R}}_s}} \right)\\
= \mathop {\max }\limits_{{{\bm{R}}_s}} \mathop {{\rm{min}}}\limits_{\bm{t}} {\rm{ trace}}\left( {{\bm{G}}{{({\bm{t}})}^{\rm{H}}}{\bm{R}}_c^{ - 1}{\bm{G}}({\bm{t}}){{\bm{R}}_s}} \right)
\end{array},
\label{MinimaxEqualitySDP}
\end{equation}
where $\bm{R}_s$ and $\bm{t}$ belong to the feasible set of $\tilde {\cal P}_r^P$. 

Combining (\ref{RadarPlayerSDP}), (\ref{TargetPlayerSDP})
and (\ref{MinimaxEqualitySDP}), we know that the Nash equilibrium exists for the game modeled by $\tilde {\cal P}_r^C$ and $\tilde {\cal P}_t^C$. Further, both $\tilde {\cal P}_r^C$ and $\tilde {\cal P}_t^C$ belong to the convex-concave minimax problems, and the Nash equilibrium can be approximately solved by the following iterative first order method\cite{nouiehed2019solving}. In particular, \vspace{2pt} we start the algorithm with an initial point ($\bm{R}_s^{(0)}$,$\bm{t}^{(0)}$)  and let 
\begin{equation}
{\tilde z^C}\left( {{{\bm{R}}_s},{\bm{t}}} \right) = {z^C}\left( {{{\bm{R}}_s},{\bm{t}}} \right) - \beta \left\| {{{\bm{R}}_s} - {\bm{R}}_s^{(0)}} \right\|_F^2,
\end{equation}
where $\beta \ge 0$ is a proximal parameter to make sure ${\tilde z^C}\left( {{{\bm{R}}_s},{\bm{t}}} \right)$ is strongly concave with respect to ${\bm{R}}_s$. Then, we carry out the following iteration\cite{wang2020improved,lin2020near,nouiehed2019solving,razaviyayn2020nonconvex} 
\begin{equation}
\left\{ {\begin{array}{*{20}{l}}
	{{\bm{R}}_s^{(t + 1)} = \arg \mathop {\max }\limits_{\left\{ {{{\bm{R}}_s}|{{\bm{R}}_s}(i,i) = \frac{{{e_t}}}{{{N_T}L}},{{\bm{R}}_s} \succeq 0} \right\}} {\tilde{z}^C}\left( {{{\bm{R}}_s},{{\bm{t}}^{(t)}}} \right)}\\
	{{\bm{t}}^{(t + 1)} = {{\rm{Proj}_\Omega}\left( {{\bm{t}}^{(t)} - \eta {\nabla _{\bm{t}}}{{\tilde z}^C}\left( {{\bm{R}}_s^{(t + 1)},{{\bm{t}}^{(t)}}} \right)} \right)}}
	\end{array}} \right.,
\label{ABRCMSC}
\end{equation}
until the gap 
\begin{equation}
{g^C}(t,t + 1) = \left| {{{\tilde z}^C}\left( {{\bm{R}}_s^{\left( t \right)},{{\bm{t}}^{(t)}}} \right) - {{\tilde z}^C}\left( {{\bm{R}}_s^{(t + 1)},{{\bm{t}}^{(t + 1)}}} \right)} \right|
\label{ABRGap}
\end{equation}
is small enough, where $\rm{Proj}_\Omega(\cdot)$ denotes the projection on $\Omega$ and $\eta$ is the iteration step size. 

Note that ${\tilde z^C}\left( {{{\bm{R}}_s},{\bm{t}}} \right)$ is strongly concave, thus $\bm{R}_s^{(t+1)}$ can be solved by CVX in polynomial time. In addition, since $\Omega$ is a scaled sphere centered by $\bm{t}_0$, $\rm{Proj}_\Omega(\bm{t})$ is given by
\begin{equation}
{\rm{Pro}}{{\rm{j}}_\Omega }\left( {\bm{t}} \right) = \left\{ \begin{array}{l}
{\bm{t}},\;{\bm{t}} \in \Omega \\
{{\bm{t}}_0} + \frac{{r\left( {{\bm{t}} - {{\bm{t}}_0}} \right)}}{{{{\left\| {{\bm{t}} - {{\bm{t}}_0}} \right\|}_2}}},\;{\bm{t}} \notin \Omega 
\end{array} \right. . 
\end{equation}

%
%
%
%
%
%

Once the equilibrium strategy $\left( {{\bm{R}}_{s,C}^*,{\bm{t}}_C^*} \right)$ is obtained, the following task is to synthesize practical waveform-filter pair $\left( {{\bm{s}}_C^*,{\bm{w}}_C^*} \right)$ complying with CM-SC from $\left( {{\bm{R}}_{s,C}^*,{\bm{t}}_C^*} \right)$. 

An inspection of the similarity constraint reveals that the phase of constant modulus signal needs to meet the following condition 
\begin{equation}
\arg {\bm{s}}(i) \in \left[ {\arg {{\bm{s}}_0}(i) - \varphi ,\arg {{\bm{s}}_0}(i) + \varphi } \right]
\end{equation} 
with $\varphi  = \rm{accos}\left( {1 - \frac{{{\delta ^2}}}{2}} \right)$.

In order to get the feasible $\bm{s}$ of  ${\cal P}_r^C$, the randomization schemes\cite{cui2013mimo} are used. More precisely, we generate $M$ random vectors $\bm{\xi}_m$ from ${\cal{CN}}\left( {{\bf{0}},{\bm{R}}_{s,C}^* \odot \left( {{{{\bm{\bar s}}}_0}{\bm{s}}_0^{\rm{T}}} \right)} \right)$, and adjust its phase with 
\begin{equation}
{{\bm{s}}^{(m)}}(i) = {{\bm{s}}_0}(i){e^{\frac{{j\left( {\arg ({{\bm{\xi }}_m}(i)) - \pi } \right)\varphi }}{{2\pi }}}}.
\label{SCWaveform}
\end{equation} 
Thus, $\bm{s}^{(m)}$ meets CM-SC due to $\frac{{\left( {\arg ({{\bm{\xi }}_m}(i)) - \pi } \right)\varphi }}{{2\pi }} \in \left[ { - \varphi ,\varphi } \right)$. Finally, the optimal waveform is selected from $\left\{ {{{\bf{s}}^{(m)}}} \right\}_{m = 1}^M$ as the one who performs best. In more detail, denote by $\bm{t}_C^{(m)}$ and ${ {{\rm{SINR}}}^{(m)}}$ the optimal sultion and optimal value for the following problem 
\begin{equation}
\begin{array}{l}
\mathop {{\rm{min}}}\limits_{\bm{t}} {{\bm{t}}^{\rm{H}}}{\left( {{\bm{H}}({{\bm{s}}^{(m)}})} \right)^{\rm{H}}}{\bm{R}}_c^{ - 1}{\bm{H}}({{\bm{s}}^{(m)}}){\bm{t}}\\
s.t.\;\;\;{\left\| {{\bm{t}} - {{\bm{t}}_0}} \right\|_2} \le r
\end{array}
\label{WorstcaseTIR}
\end{equation}
respectively, and we pick the optimal signal ${{\bm{s}}^{(k)}}$ achieving the maximum value ${{{\rm{SINR}}}^{(k)}}$. To this end, we get the optimal waveform-filter pair with
\begin{equation}
{\bm{s}}_C^* = {{\bm{s}}^{(k)}},{\bm{w}}_C^* = {\bm{R}}_c^{ - 1}{\bm{H}}({\bm{s}}_C^*){\bm{t}}_C^{(k)}.
\label{OptimalWaveformFilterCMSC}
\end{equation}

The overall procedure for waveform-filter design under CM-SC is summarized in Table \ref{Algorithm2CMSC} as \textbf{Algorithm 2}. The convergence of \textbf{Algorithm 2} is determined by the iteration steps described by (\ref{ABRCMSC}). In fact, it converges if the step size $\eta$ is properly chosen and the choice of $\eta$ can be found in \cite{nouiehed2019solving}. However, it goes beyond the scope of our paper. The computational complexities consist of two parts, namely, iteration algorithm and waveform-filter pair synthesis. One can see that the computational complexities of the iteration algorithm is proportion to the number of iterations. At each iteration, it requires at most $O\left( {{{\left( {L{N_T}} \right)}^{9}}} \right)$\cite{NemirovskiLectures} 
to optimize $\bm{R}_s^{(t+1)}$ and $O\left( {{{{Q}}^{2}}} \right)$ to compute $\bm{t}^{(t+1)}$. As to the synthesis stage, it requires $O\left( {{M\left( {\left( {L{N_T}} \right)}^{2}+Q^3\right)}}\right)$\cite{li2003robust,aubry2012cognitive} to generate $\left\{ {{{\rm{SINR}}^{(m)}}} \right\}_{m = 1}^M$ and $O\left( {\left( {L{N_R} + Q{N_R}} \right)^2}\right) $ to calculate $\bm{w}_C^{*}$.  Therefore, the total \vspace{2pt}computational complexities are dominated by its highest order $O\left( {{{\left( {L{N_T}} \right)}^{9}}} \right)$.

\begin{table}[!t]
	\renewcommand{\arraystretch}{1.3}
	\caption{Algorithm for waveform-filter optimization with CM-SC}
	\label{Algorithm2CMSC}
	\centering
	\begin{tabular}{l}
		\hline
		\textbf{Input}:
		$\bm{a}(\theta_{t})$, $\bm{b}(\theta_{t})$, $\bm{s}_0$, $e_t$, $\delta$,
		${{\bm{R}}_{c}}$, ${{\bm{t}}_{0}}$, $r$, $\beta$, $\eta$, $\epsilon^C$ and $M$.\\ 
		\hline 
		\@ \@\@ \@ \textit{Step 1}: Initialize $t=0$,  $\bm{t}^{(0)} = \bm{t}_0$ and $\bm{R}_s^{(0)} = \bm{s}_0\bm{s}_0^{\rm{H}}$. \\
		\@ \@\@ \@ \textit{Step 2}: Solve optimization problems described in (\ref{ABRCMSC}) alternatively. \\
		\@ \@\@ \@ \textit{Step 3}: Verify inequality $g^C(t,t+1) \le \epsilon^C$. If true, ${\bm{R}}_{s,C}^* = {\bm{R}}_{s}^{(t+1)}$, \\
		\@ \@\@ \@ \@ \@\@ \@ \@ \@\@ \@ \@ \@\@ \@ \@ ${\bm{t}}_C^*={\bm{t}}^{(t+1)}$  and go to  \textit{Step 4}, otherwise, $t=t+1$ and go\\ 
		\@ \@\@ \@ \@ \@\@ \@ \@ \@\@ \@ \@ \@\@ \@ \@ to \textit{Step 2}.\\
		\@ \@\@ \@ \textit{Step 4}: Generate $\left\{ {{{\bm{\xi }}_{m}}} \right\}_{m = 1}^M$ from ${\cal{CN}}\left( {0,{\bm{R}}_{s,C}^* \odot \left( {{{{\bm{\bar s}}}_0}{\bm{s}}_0^{\rm{T}}} \right)} \right)$ and \\
		\@ \@\@ \@ \@ \@\@ \@ \@ \@\@ \@ \@ \@\@ \@ \@  calculate  $\left\{ {{{\bm{s}}^{(m)}}} \right\}_{m = 1}^M$ with (\ref{SCWaveform}). \vspace{2pt}\\		
		\@ \@\@ \@ \textit{Step 5}: For each  ${{\bm{s}}^{(m)}}$, optimize (\ref{WorstcaseTIR}) and record the corresponding\\
		\@ \@\@ \@ \@ \@\@ \@ \@ \@\@ \@ \@ \@\@ \@ \@  optimal value ${\rm{SINR}}^{(m)}$ and solution $\bm{t}_C^{(m)}$.\vspace{2pt}\\
		\@ \@\@ \@ \textit{Step 6}: Pick the maximal value in $\left\{ {{{\rm{SINR}}^{(m)}}} \right\}_{m = 1}^M$, for example,\\
		\@ \@\@ \@ \@ \@\@ \@ \@ \@\@ \@ \@ \@\@ \@ \@ ${\rm{SINR}}^{(k)}$. Finally, synthesize $\bm{s}_C^{*}$ and $\bm{w}_C^{*}$ with (\ref{OptimalWaveformFilterCMSC}).\\
		\hline
		\textbf{Output}: 
		\@ \@ ${\bm{s}_C^{*}}$ and ${\bm{w}_C^{*}}$.\\
		\hline
	\end{tabular}
\end{table}

\subsection{SC-SC on waveform}
This subsection is devoted to the Stackelberg equilibrium of ${\cal{P}}_r$ with $\Psi = \Psi_S$. Leveraging on the MM tools\cite{wu2017transmit}, we construct a sequence of Stackelberg games\cite{song2011mimo}, which belongs to the convex non-concave minimax problems. Further, we prove that the Stackelberg games can be equivalently solved by optimizing a convex problem. Thus, the optimal waveform can be obtained without solving the target strategies in ${\cal{P}}_t$.  

Similarly to the former cases, we recast ${\cal{P}}_r$ as follows 
\begin{equation}
{\cal{P}}_r^S\left\{ {\begin{array}{*{20}{l}}
{\mathop {{\rm{max}}}\limits_{\bm{s}} \mathop {{\rm{min}}}\limits_{\bm{t}} {{\bm{s}}^{\rm{H}}}{\bm{G}}{{({\bm{t}})}^{\rm{H}}}{\bm{R}}_c^{ - 1}{\bm{G}}({\bm{t}}){\bm{s}}}\\
{s.t.\;\;\;{\kern 1pt} \left\| {\bm{s}} \right\|_2^2 \le {e_t},{{\left\| {{\bm{s}} - {{\bm{s}}_0}} \right\|}_\infty } \le \frac{{\delta \sqrt {{e_t}} }}{{\sqrt {{N_T}L} }}}\\
{\;\;\;{\kern 1pt} \;\;\;{\kern 1pt} {{\bm{s}}^{\rm{H}}}{{\bm{R}}_I}{\bm{s}} \le {e_I},{{\left\| {{\bm{t}} - {{\bm{t}}_0}} \right\|}_2} \le r}
\end{array}} \right..
\end{equation}

Note that the feasibility of ${\cal{P}}_r^S$ is well discussed in \cite{aubry2015new,aubry2016optimization}, and it is beyond the scope of our paper, where $e_I$ is carefully set to make sure that ${\cal{P}}_r^S$ is always feasible. 

The key point of MM algorithm for solving the maximization problem is to find a proper minorizer of objective. And the following proposition provides a minorizer for solving ${\cal{P}}_r^S$.  
  
\begin{propositions}
Let 
\begin{equation*}
{z^S}({\bm{s}}) = \mathop {{\rm{min}}}\limits_{{{\left\| {{\bm{t}} - {{\bm{t}}_0}} \right\|}_2} \le r}  {{\bm{s}}^{\rm{H}}}{\bm{G}}{({\bm{t}})^{\rm{H}}}{\bm{R}}_c^{ - 1}{\bm{G}}({\bm{t}}){\bm{s}},
\end{equation*}
then 
\begin{equation*}
\begin{array}{l}
{{\tilde z}^S}({\bm{s}},{{\bm{s}}^{(l)}}) = \mathop {{\rm{min}}}\limits_{{{\left\| {{\bm{t}} - {{\bm{t}}_0}} \right\|}_2} \le r}  2{\mathop{\rm Re}\nolimits} \left\{ {{{\left( {{{\bm{s}}^{(l)}}} \right)}^{\rm{H}}}{\bm{G}}{{({\bm{t}})}^{\rm{H}}}{\bm{R}}_c^{ - 1}{\bm{G}}({\bm{t}}){\bm{s}}} \right\} - \\
\quad \quad \quad \quad \quad \quad \quad \quad {\left( {{{\bm{s}}^{(l)}}} \right)^{\rm{H}}}{\bm{G}}{({\bm{t}})^{\rm{H}}}{\bm{R}}_c^{ - 1}{\bm{G}}({\bm{t}}){{\bm{s}}^{(l)}}
\end{array}
\end{equation*} 
is a minorizer of ${z^S}({\bm{s}})$ at $\bm{s}^{(l)}$.
\end{propositions}
\begin{IEEEproof}
See Appendix C.
\end{IEEEproof}

Leveraging on the MM algorithm, the Stackelberg equilibrium described by ${\cal{P}}_r^S$ can be solved by sequentially optimizing $\left\{ {{\cal{P}}_r^{S,(l)}} \right\}_{l = 1}^\infty$ with 
\begin{equation}
{\cal{P}}_r^{S,(l)}\left\{ {\begin{array}{*{20}{l}}
	{{{\bm{s}}^{(l + 1)}} = \arg \mathop {{\rm{max}}}\limits_{\bm{s}} {{\tilde z}^S}({\bm{s}},{{\bm{s}}^{(l)}})}\\
	{s.t.\;\;\;{\kern 1pt} \left\| {\bm{s}} \right\|_2^2 \le {e_t},{{\left\| {{\bm{s}} - {{\bm{s}}_0}} \right\|}_\infty } \le \frac{{\delta \sqrt {{e_t}} }}{{\sqrt {{N_T}L} }}}\\
	{\;\;\;{\kern 1pt} \;\;\;{\kern 1pt} {{\bm{s}}^{\rm{H}}}{{\bm{R}}_I}{\bm{s}} \le {e_I}}
	\end{array}} \right.,
\end{equation}
which is also a Stackelberg game but easier to solve. 

Now, we turn to solving ${\cal{P}}_r^{S,(l)}$. It is worth pointing out that the objective ${{\tilde z}^S}({\bm{s}},{{\bm{s}}^{(l)}})$ is a linear function with respect to $\bm{s}$. Substituting (\ref{SingalModelVectorFormT}) into ${{\tilde z}^S}({\bm{s}},{{\bm{s}}^{(l)}})$, and after some algebraic manipulations, we obtain another form of it, given by 
\begin{equation}
{\tilde z^S}({\bm{s}},{{\bm{s}}^{(l)}}) = \mathop {{\rm{min}}}\limits_{\left\{ {{\bm{t}}|{{\left\| {{\bm{t}} - {{\bm{t}}_0}} \right\|}_2} \le r} \right\}} {{\bm{t}}^{\rm{H}}}{\bm{U}}({\bm{s}},{{\bm{s}}^{(l)}}){\bm{t}}
\label{ZSt}
\end{equation}
with 
\begin{equation}
\begin{array}{l}
{\bm{U}}({\bm{s}},{{\bm{s}}^{(l)}}) = {\bm{H}}{({{\bm{s}}^{(l)}})^{\rm{H}}}{\bm{R}}_c^{ - 1}{\bm{H}}{({\bm{s}})} + {\bm{H}}{({\bm{s}})^{\rm{H}}}{\bm{R}}_c^{ - 1}{\bm{H}}{({{\bm{s}}^{(l)}})}  \\
\quad \quad \quad \quad \quad\;-{\bm{H}}{({{\bm{s}}^{(l)}})^{\rm{H}}}{\bm{R}}_c^{ - 1}{\bm{H}}{({{\bm{s}}^{(l)}})}.
\end{array}
\end{equation}
Unfortunately, ${{\bm{t}}^{\rm{H}}}{\bm{U}}({\bm{s}},{{\bm{s}}^{(l)}}){\bm{t}}$ is not convex with respect to $\bm{t}$ due to ${\bm{U}}({\bm{s}},{{\bm{s}}^{(l)}}) \nsucceq 0$, even though it is a quadratic form. Thus, the iteration algorithm mentioned in  (\ref{ABRCMSC}) is invalid when solving ${\cal{P}}_r^{S,(l)}$. To this end, we have to devise another algorithm to address  ${\cal{P}}_r^{S,(l)}$, which is based on the following proposition. 

\begin{propositions}
${\cal{P}}_r^{S,(l)}$ is equivalent to the following problem 
\begin{equation*}
\hat {\cal{P}}_r^{S,(l)}\left\{ \begin{array}{l}
\mathop {\max }\limits_{{\bm{s}},\lambda ,\gamma } \;\gamma \\
s.t.\left[ {\begin{array}{*{20}{c}}
	{{\bm{U}}({\bm{s}},{{\bm{s}}^{(l)}}) + \lambda {\bf{I}}}&{\lambda {{\bm{t}}_0}}\\
	{\lambda {\bm{t}}_0^{\rm{H}}}&{\lambda {\bm{t}}_0^{\rm{H}}{{\bm{t}}_0} - \lambda {r^2} - \gamma }
	\end{array}} \right] \succeq 0 \vspace{2pt} \\
\quad\quad {\left\| {{\bm{s}} - {{\bm{s}}_0}} \right\|_\infty } \le \frac{{\delta \sqrt {{e_t}} }}{{\sqrt {{N_T}L} }},{\kern 1pt} \left\| {\bm{s}} \right\|_2^2 \le {e_t}\vspace{2pt}\\
\quad\quad {{\bm{s}}^{\rm{H}}}{{\bm{R}}_I}{\bm{s}} \le {e_I},\lambda  \ge 0
\end{array} \right.,
\end{equation*}
where $\lambda$ and $\gamma$ are auxiliary variables.
\end{propositions}
\begin{IEEEproof}
	See Appendix D.\vspace{2pt}
\end{IEEEproof}
 
Obviously, $\hat {\cal{P}}_r^{S,(l)}$ is a SDP problem since ${\bm{U}}({\bm{s}},{{\bm{s}}^{(l)}})$ is linear with respect to $\bm{s}$. Based on \textit{Proposition 3}, we sequentially solve ${\cal{P}}_r^{S,(l)}$ until converges, and get the optimal waveform $\bm{s}_S^*$.
	
Once the optimal waveform $\bm{s}_S^*$ is obtained, we need to calculate the optimal filter $\bm{w}_S^*$ to construct the Stackelberg equilibrium strategy of radar. In oder to achieve this goal, we first get the target player's strategy in the Stackelberg game by solving 
\begin{equation}
\begin{array}{l}
{\bm{t}}_{{S}}^{\bm{*}} = \arg \min {{\bm{t}}^{\rm{H}}}{\bm{H}}{({\bm{s}}_S^*)^{\rm{H}}}{\bm{R}}_c^{ - 1}{\bm{H}}({\bm{s}}_S^*){\bm{t}}\\
s.t.\quad \left\| {{\bm{t}} - {{\bm{t}}_0}} \right\|_2^{} \le r
\end{array},
\label{OptimalTIRSpectral}
\end{equation}
then we calculate $\bm{w}_S^*$ with
\begin{equation}
{\bm{w}}_S^* = {\bm{R}}_c^{ - 1}{\bm{H}}({\bm{s}}_S^*){\bm{t}}_S^{*}.
\label{OptimalFilterSCSC}
\end{equation}

The overall procedure for waveform-filter design under SC-SC is summarized in Table \ref{Algorithm3SCSC} as \textbf{Algorithm 3}.
The convergence of \textbf{Algorithm 3} is ensured by the MM algorithm\cite{hunter2004tutorial} owing to the fact that ${z^S}({\bm{s}})$ is upper bounded by $e_t{\lambda _{\max }}\left( {{\bm{G}}{{({{\bm{t}}_0})}^H}{\bm{R}}_c^{ - 1}{\bm{G}}({{\bm{t}}_0})} \right)$. As to the computational complexities, it is proportional to the number of iterations of MM algorithm when optimizing $\bm{s}_S^{*}$. At each iteration, it requires at most $O\left( {{{\left( {Q + L{N_T}} \right)}^{0.5}}\left( {{{\left( {L{N_T}} \right)}^4} + {{\left( {QL{N_T}} \right)}^2} + {Q^3}L{N_T}} \right)} \right)$. Additionally, it requires at most $O\left( {{Q^3} + {{\left( {L{N_R} + Q{N_R}} \right)}^2}} \right)$ to calculate $\bm{w}_S^{*}$. Therefore, the total computational complexities are dominated by the highest order  $O\left( {{{\left( {Q + L{N_T}} \right)}^{0.5}}\left( {{{\left( {L{N_T}} \right)}^4} + {{\left( {QL{N_T}} \right)}^2} + {Q^3}L{N_T}} \right)} \right)$.

\begin{table}[!t]
	\renewcommand{\arraystretch}{1.3}
	\caption{Algorithm for waveform-filter optimization with CM-SC}
	\label{Algorithm3SCSC}
	\centering
	\begin{tabular}{l}
		\hline
		\textbf{Input}:
		$\bm{a}(\theta_{t})$, $\bm{b}(\theta_{t})$, $\bm{s}_0$, $e_t$, $\delta$,
		${{\bm{R}}_{c}}$, ${{\bm{t}}_{0}}$, $r$, $[f_1^k,f_2^k]$, $\alpha_k$, $e_I$ and $\epsilon^S$.\\ 
		\hline
		\@ \@\@ \@ \textit{Step 1}: Initialize $l=0$ and $\bm{s}^{(0)}$ with any feasible waveform;\\
		\@ \@\@ \@ \@ \@\@ \@ \@ \@\@ \@ \@ \@\@ \@ \@  let $\tilde{z}^S(\bm{s}^{(0)},\bm{s}^{(-1)}) = {z}^S(\bm{s}^{(0)})$.  \\
		\@ \@\@ \@ \textit{Step 2}: Solve optimization problem ${\cal{P}}_r^{S,(l)}$.\\
		\@ \@\@ \@ \textit{Step 3}: Verify   $\tilde{z}^S(\bm{s}^{(l+1)},\bm{s}^{(l)})-\tilde{z}^S(\bm{s}^{(l)},\bm{s}^{(l-1)}) \le \epsilon^S$.\\
		 \@ \@\@ \@ \@ \@\@ \@ \@ \@\@ \@ \@ \@\@ \@ \@ If true, 
		go to  \textit{Step 4}, otherwise, $l=l+1$ and go to \textit{Step 2}.\\ 
		\@ \@\@ \@ \textit{Step 4}: Let $\bm{s}_S^{*} = \bm{s}^{(l+1)}$; solve (\ref{OptimalTIRSpectral}) and obtain $\bm{w}_S^{*}$ with (\ref{OptimalFilterSCSC}).\\
		\hline
		\textbf{Output}: 
		\@ \@ ${\bm{s}_S^{*}}$ and ${\bm{w}_S^{*}}$.\\
		\hline
	\end{tabular}
\end{table}

\section{Numerical experiments}
In this section, several numerical experiments are conducted 
to assess the performance of the proposed algorithms. Unless otherwise specified, in the following experiments, we assume a
colocated MIMO radar system with $N_T= 2$ transmitters
and $N_R = 4$ receivers, where the inter-element space is wave-length for transmitters and half wave-length for receivers.
The carrier frequency is 3GHz, and the code length is $L = 16$. 
Meanwhile, the target T is assumed at $\theta_{t} = $30$^\circ$ azimuth with a prescribed
\begin{equation*}
{{\bm{t}}_{\rm{0}}}{\rm{ = }}{\left[ {{\rm{0}}{\rm{.2}}{e^{\frac{{j\pi }}{4}}}{\rm{,0}}{\rm{.3}}{e^{\frac{{j\pi }}{3}}}{\rm{,0}}{\rm{.8,0}}{\rm{.3}}{e^{ - \frac{{j\pi }}{6}}}{\rm{,0}}{\rm{.2}}{e^{ - \frac{{j\pi }}{3}}},0.1{e^{ - \frac{{j\pi }}{3}}}} \right]^{\rm{T}}}\in \mathbb{Q}^{6}.
\end{equation*}
As to the noise, we also assume that it is complex Gaussian distribution with ${\bm{n}}\sim{\cal{CN}}\left( {{\bf{0}},{{\bm{R}}_n}} \right)$, where ${{\bm{R}}_n}\left( {m,n} \right) = {0.8^{  \left| {m - n} \right|}}$.
All numerical experiments are analyzed using Matlab
2014a version and performed in a standard PC (with CPU Core
i5 3.0 GHz and 16 GB RAM).

\subsection{Experiments for EC on waveform}
Now, we consider the algorithm for EC on waveform, namely, \textbf{Algorithm 1}. Given that the algorithm in \cite{chen2009mimo} are proposed to address the EC on waveform, we also give comparisons with it in terms of detection probability and running time. 

Fig.\ref{DetectionProbabilityCurve} depicts the detection probability versus the transmit energy $e_t$ for different $r$ by substituting $\underline {{\rm{SINR}}}_E$ or $\overline {{\rm{SINR}}}_E$ into (\ref{MarcumQFunction}) with $P_{fa} = 10^{-6}$. As expected, the detection probability monotonically increases with respect to $e_t$ for a fixed $r$. Meanwhile, it monotonically decreases with respect to $r$ for a fixed $e_t$, since a larger $r$ means a "smarter" target, which is unfavorable from the view of radar player. Moreover, one can observe from the figure that \textbf{Algorithm 1} achieves almost the same results as the algorithm in \cite{chen2009mimo} does, which demonstrates the effectiveness of the proposed \textbf{Algorithm 1}. Next we compare the running time of the two algorithms, and illustrate the results in Table \ref{RunningTime}. According to the data in Table \ref{RunningTime}, we find that less running time is needed for \textbf{Algorithm 1} compared with its counterpart in \cite{chen2009mimo}. The reason can be explained that \textbf{Algorithm 1} solves the optimal waveform-filter directly based on \textit{Theorem 1} without outer iteration, thus only the inner iteration is needed for optimizing $\hat{{\cal{P}}}_t^E$. By contrast, the running time of algorithm in \cite{chen2009mimo} depends on not only inner iteration of optimization $\bm{t}$, but also the number of outer iterations. 

Fig.\ref{ModulusWaveform} illustrates modulus of the optimal waveform with $r=0.8$ and $e_t =1$ for two algorithms. One can see serious fluctuation of transmit waveform, which will not meet the demands of radar transmitter given that the transmitter always operates at saturation situation. These results inspire us to find more suitable waveform to fulfill the detection task.

\begin{table}[!t]
	\renewcommand{\arraystretch}{1.3}
	\caption{Running time for Algorithm 1 and the algorithm in \cite{chen2009mimo} with $e_t =1$}
	\label{RunningTime}
	\centering
	\begin{tabular}{ccc}
		\hline
		& \textbf{Algorithm 1} & Algorithm in \cite{chen2009mimo} \\
		\hline
		$r=0.1$ &  {2.03s}    &  5.37s\\
		$r=0.3$ &  {1.83s}    &  4.44s\\
		$r=0.5$ &  {1.83s}    &  4.35s\\
		$r=0.8$ &  {1.84s}    &  4.39s\\
		\hline
		\end{tabular}
\end{table}

\begin{figure}[!t]
	\centering
	\includegraphics[width=3.2in]{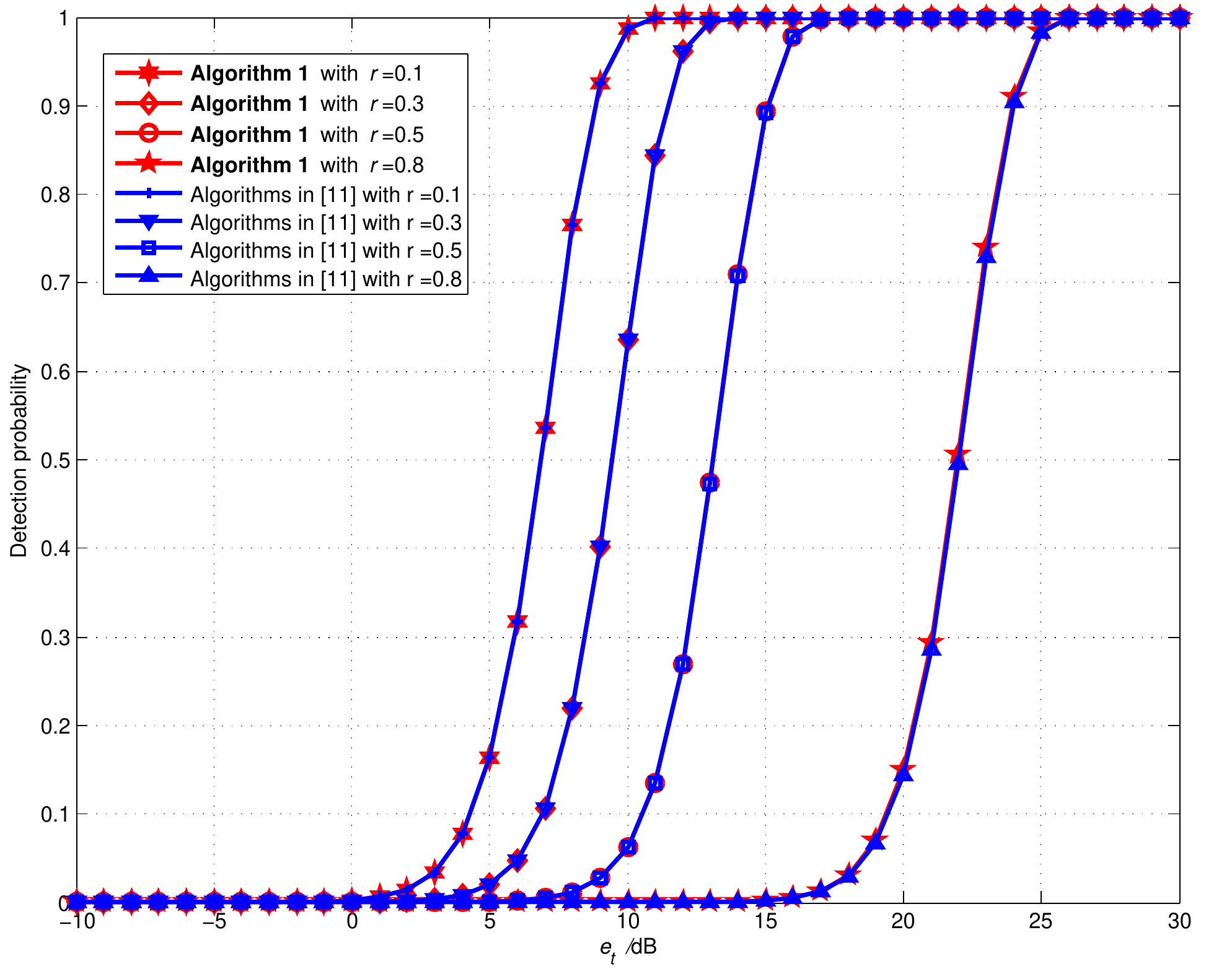}
	\caption{Detection probability versus $e_t$ for different $r$}
	\label{DetectionProbabilityCurve}
\end{figure}

\begin{figure}[!t]
	\centering
	\includegraphics[width=3.2in]{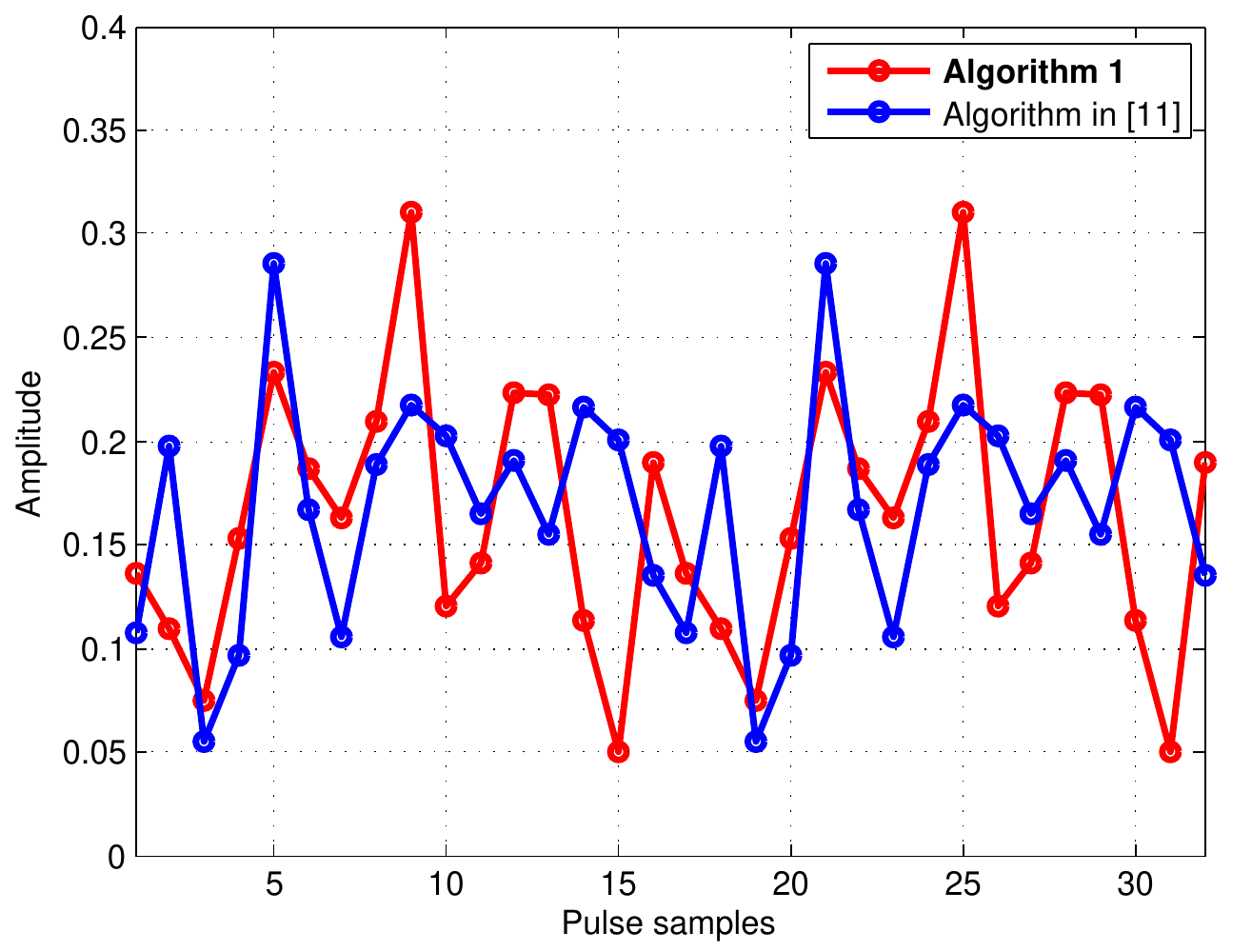}
	\caption{Modulus of waveform}
	\label{ModulusWaveform}
\end{figure}

\subsection{Experiments for CM-SC on waveform}
In this subsection, we consider the algorithm for CM-SC on waveform, namely, \textbf{Algorithm 2}. In the following experiments, the parameters $\beta$, $\eta$ and $\epsilon^C$ are set to be 0.05, 0.002 and 0.001, respectively. Moreover, we conduct $M=100$ trails when synthesizing waveform with randomization. As to the reference waveform, the orthogonal LFM is chosen with $\bm{s}_0 = {\rm{vec}}(\bm{S}_0)$, where ${{\bm{S}}_{\rm{0}}}$ is defined by\cite{cui2013mimo,cheng2017mimo,aldayel2016successive}   
\begin{equation}
{{\bm{S}}_{\rm{0}}}{\rm{(}}n{\rm{,}}l{\rm{) = }}\sqrt {\frac{{{e_t}}}{{{N_T}L}}} {e^{\frac{{j\pi (2n(l - 1) + {{(l - 1)}^2})}}{L}}}.
\end{equation}

Fig.\ref{ConvergenceCurve} depicts the value of $\tilde{z}^C(\bm{R}_s,\bm{t})$ versus the number of iterations for different $r$. An inspection of Fig.\ref{ConvergenceCurve} reveals that \textbf{Algorithm 2} converges after several iterations and number of iterations increases with respect to $r$ due to the expansive of $\Omega$. More precisely, the algorithm converges after 3 iterations for $r=0.1$, but 52 iterations for $r=0.8$.  

Next, we investigate the detection performance of the optimal waveform-filter pair for different $\delta$ by substituting $\underline {{\rm{SINR}}}_C$ into (\ref{MarcumQFunction}) with $P_{fa}=10^{-6}$, where $r$ is set to be 0.8. We also give comparisons with the algorithm in \cite{karbasi2015robust} realizing that the algorithm in \cite{karbasi2015robust} can be applied to the similarity constraint by modifying its randomization schemes. The corresponding results are shown in Fig.\ref{DetecionProbabilityCMSC}. As expected, the detection probability increases with respect to $e_t$ as well as $\delta$, which is consistent with our intuition, since a higher transmit energy means a higher SINR and a larger $\delta$ means more freedom of waveform. It is observed from the figure that our algorithm shows its superiority over the algorithm in \cite{karbasi2015robust} in terms of detection probability with the same $\delta$. 

To be honest, compared with the algorithm in \cite{karbasi2015robust}, our algorithm suffer heavier computational burden, even though it achieves a higher SINR. The relevant results are illustrated in Table \ref{RunningTimeAlgorithm2}. It can be seen that the running time of our algorithm increases with respect to $r$. We explain the reason that solving (\ref{ABRCMSC}) is time consuming at each iteration and the number of iterations is also large for $r=0.8$(see Fig.\ref{ConvergenceCurve}). While the running time of algorithm in \cite{karbasi2015robust} is relatively stable with respect to $r$, since $\Omega$ is approximated by randomly sampling. Nevertheless, it is worth pointing out that the number of iterations can be reduced if the parameters $\beta$ and $\eta$ are carefully chosen, and interested readers may refer to \cite{nouiehed2019solving,wang2020improved,lin2020near}.  

In addition, Fig.\ref{PulseCompressionCMSC} depicts the properties of pulse compression for different $\delta$. These results display that larger $\delta$ suffers from higher sidelobe levels, even though a higher detection probability is achieved. In particular, the highest sidelobe level for $\delta = 1$ is about -6dB, while it is about -13dB for $\delta = 0.1$. Moreover, the sidelobe level achieved by {\bf{Algorithm 2}} is slightly higher than its counterpart, which also demonstrates the trade-off
between better SINR and low side lobes.

\begin{figure}[!t]
	\centering
	\includegraphics[width=3.2in]{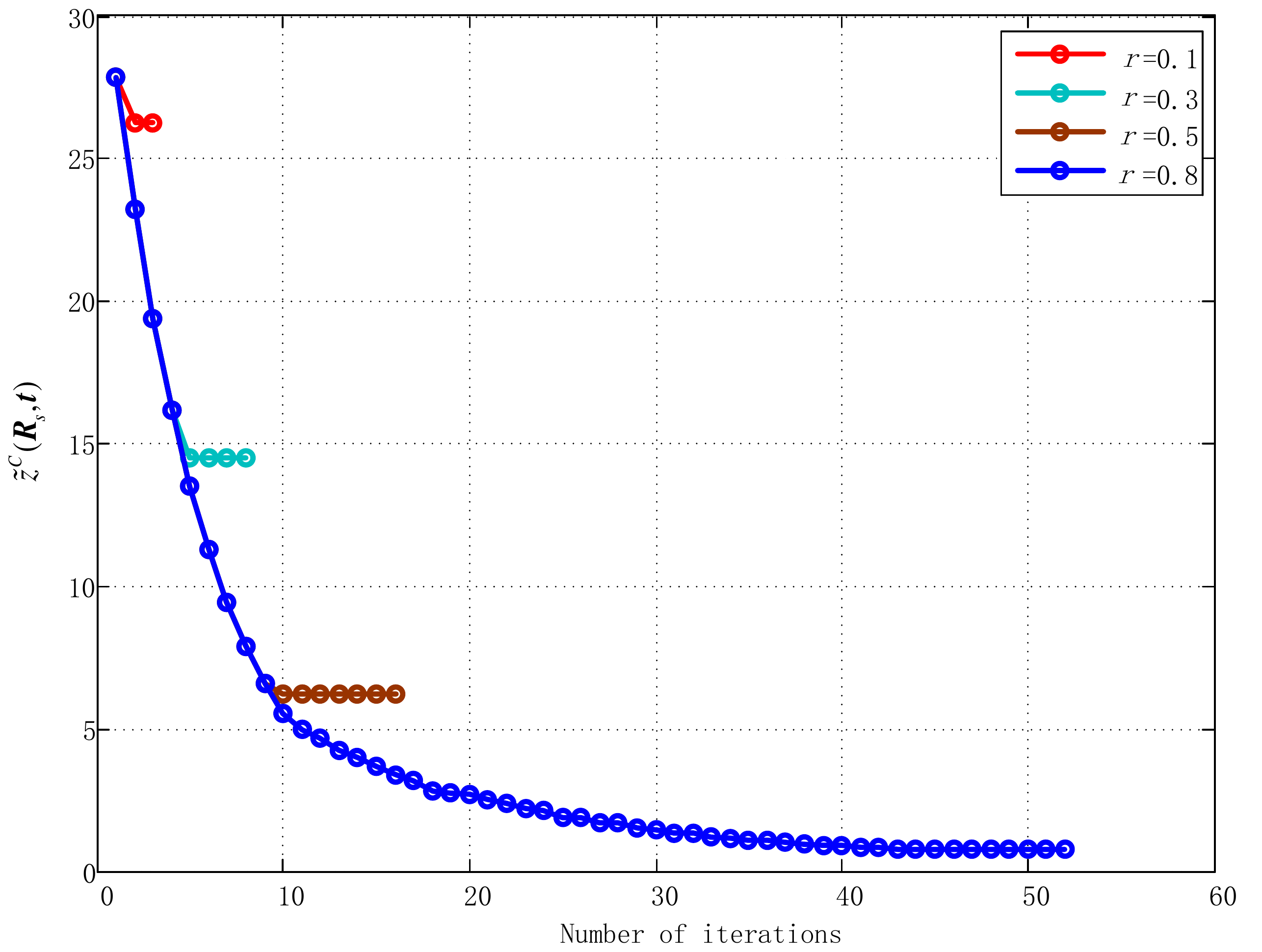}
	\caption{$\tilde{z}^C(\bm{R}_s,\bm{t})$ versus number of iterations for different $r$}
	\label{ConvergenceCurve}
\end{figure}

\begin{table}[!t]
	\renewcommand{\arraystretch}{1.3}
	\caption{Running time for Algorithm 2 and the algorithm in \cite{karbasi2015robust} with $e_t =1$ and $\delta = 1$}
	\label{RunningTimeAlgorithm2}
	\centering
	\begin{tabular}{ccc}
		\hline
		& \textbf{Algorithm 2} & Algorithm in \cite{karbasi2015robust} \\
		\hline
		$r=0.1$ &  {19.81s}    &  15.80s\\
		$r=0.3$ &  {28.74s}    &  15.83s\\
		$r=0.5$ &  {41.42s}    &  15.98s\\
		$r=0.8$ &  {110.63s}    &  16.11s\\
		\hline
	\end{tabular}
\end{table}

\begin{figure}[!t]
	\centering
	\includegraphics[width=3.2in]{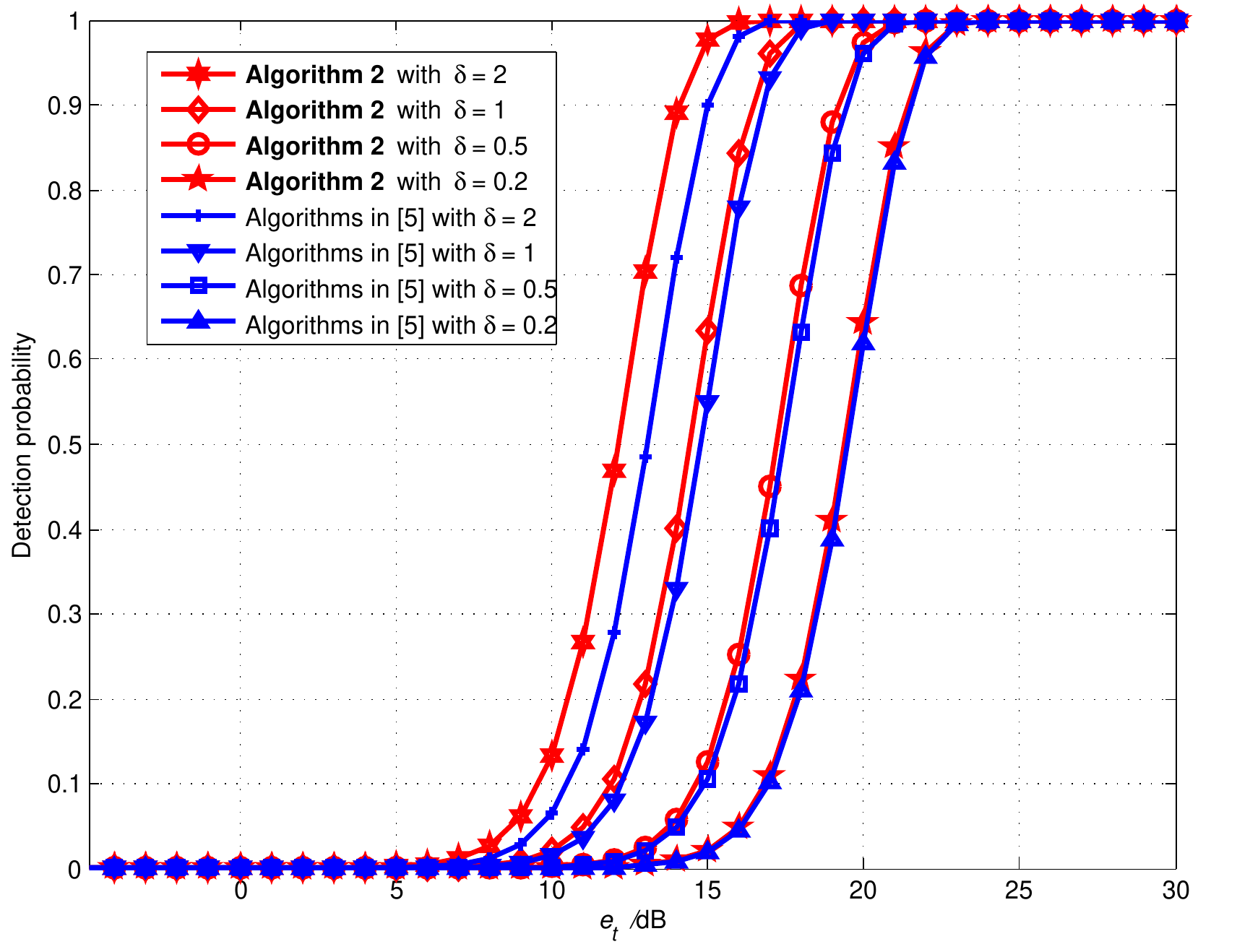}
	\caption{Detection probability versus $e_t$ for different $\delta$}
	\label{DetecionProbabilityCMSC}
\end{figure}

\begin{figure*}[!t]
	\centering
	\begin{tabular}{cc}
		\includegraphics[width=3.2in,height=2.4in]{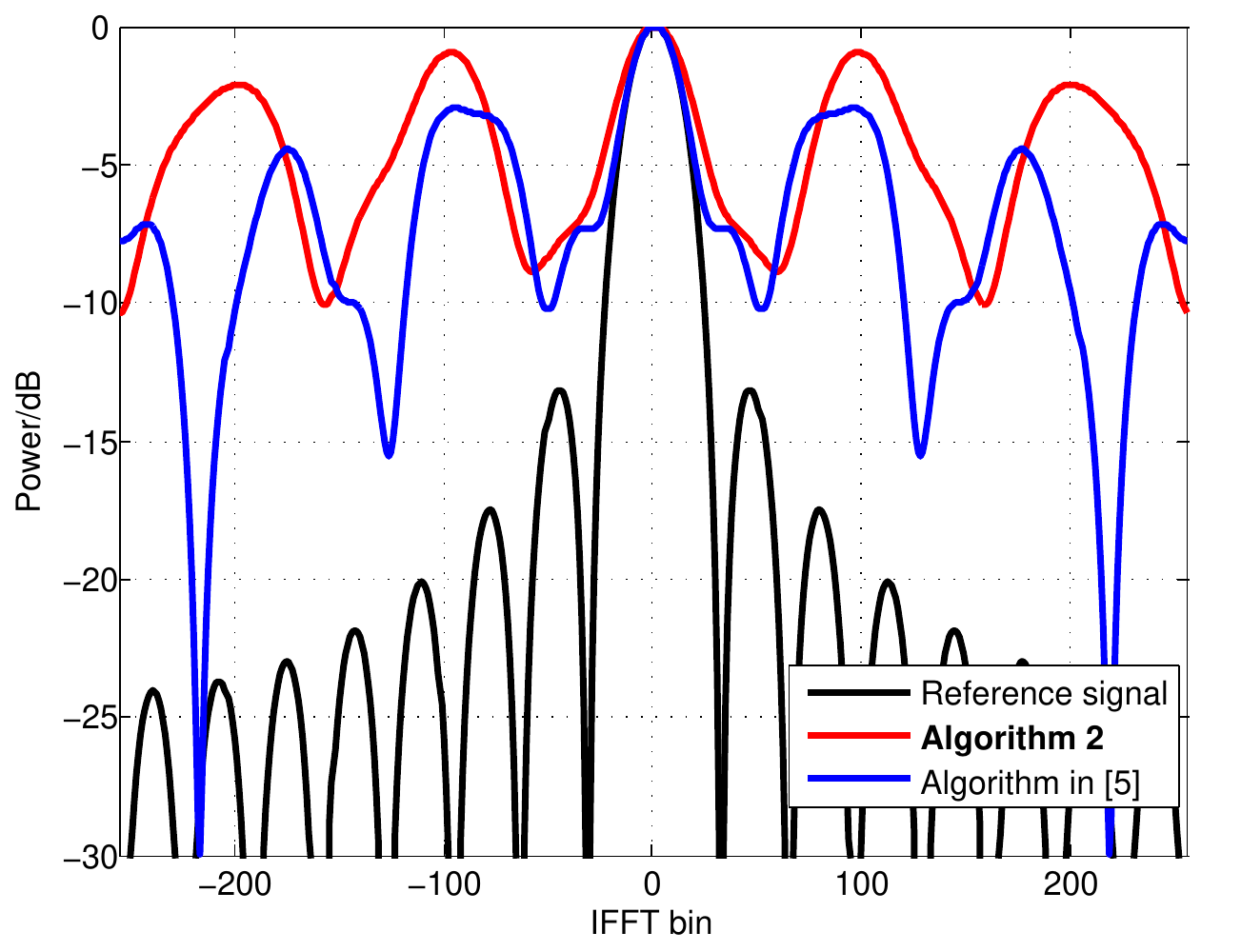}&
		\includegraphics[width=3.2in,height=2.4in]{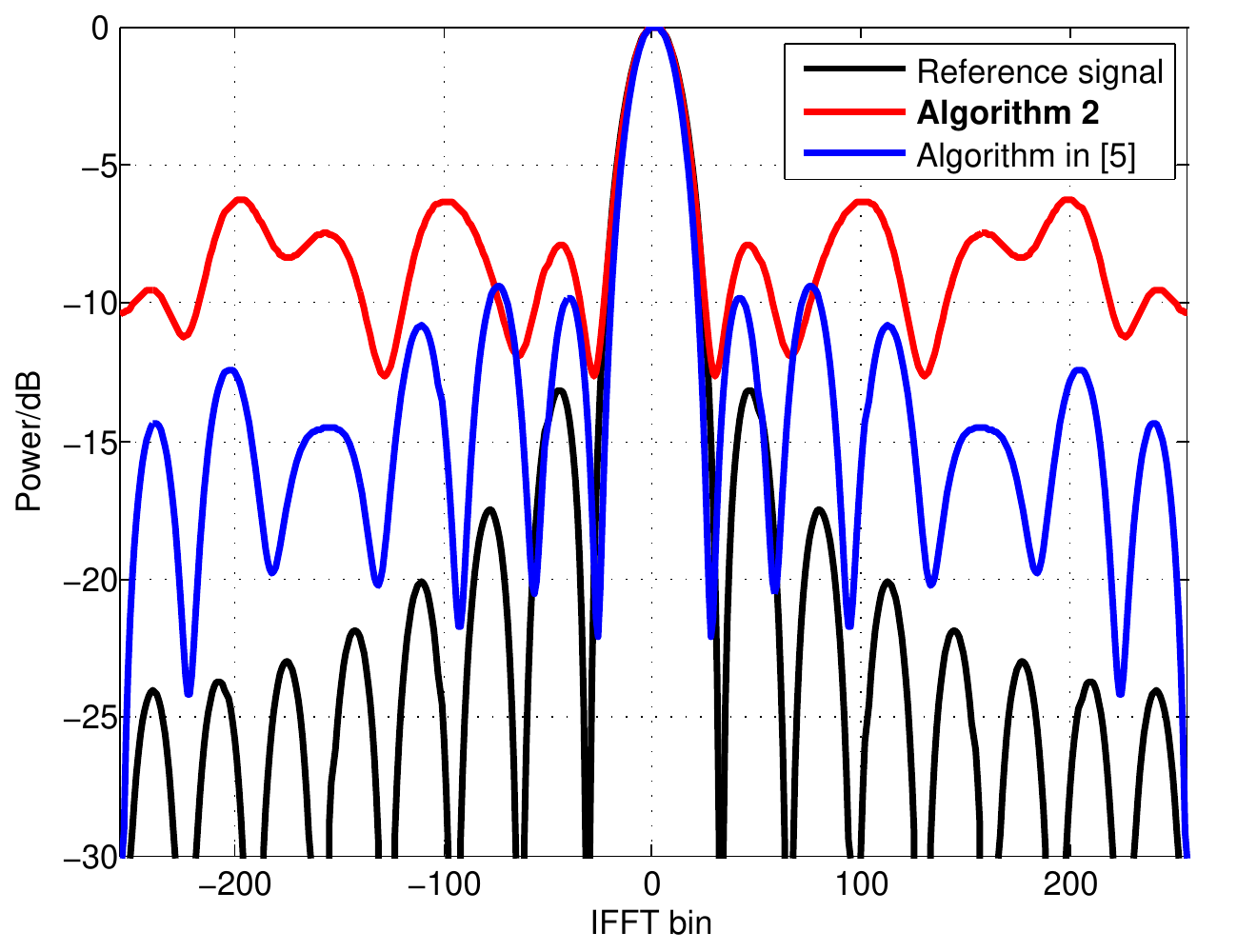}\\
		(a) & (b)\\
		\includegraphics[width=3.2in,height=2.4in]{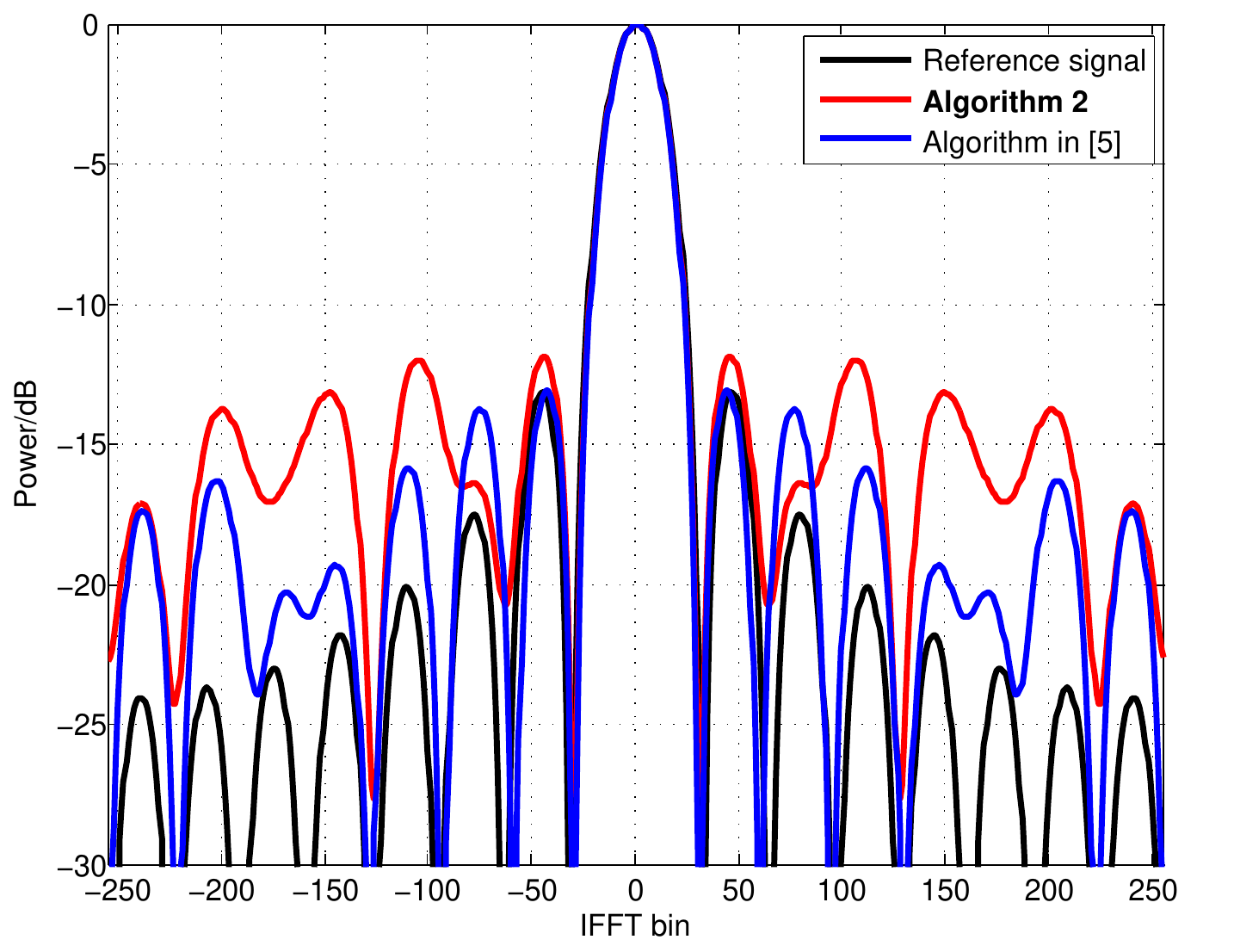}&
		\includegraphics[width=3.2in,height=2.4in]{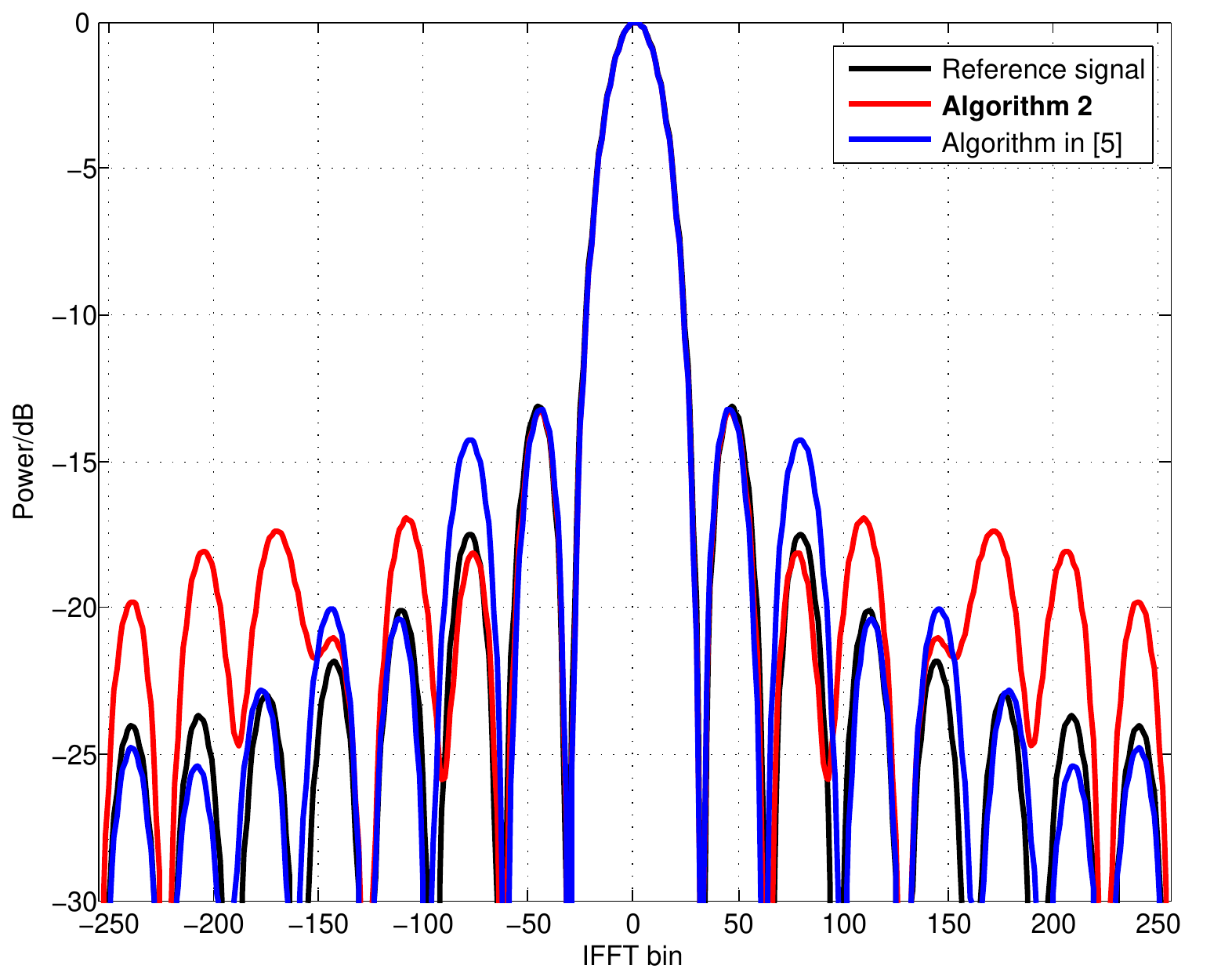}\\
		(c) & (d)\\
	\end{tabular}
	\centering
	\caption{Pulse compression for different $\delta$: (a) $\delta =2 $,  (b) $\delta =1 $, (c) $\delta =0.5 $, (d) $\delta =0.2 $}
	\label{PulseCompressionCMSC}
\end{figure*}

%

\subsection{Experiments for SC-SC on waveform}
At last, we consider the algorithm for SC-SC on waveform, namely, \textbf{Algorithm 3}. In the following experiments, the parameters $r$ and $\epsilon^S$ are set to be 0.8 and 0.001, respectively. Moreover, the orthogonal LFM waveform is also used as the reference waveform $\bm{s}_0$. As to the spectral compatibility parameters, transmission power is limited in two frequency intervals. The first one is $[f_1^1,f_2^2] =[0.30,0.40]$ with $\alpha_1 = 0.6$, and the second one is $[f_1^2,f_2^2] =[0.60,0.80]$ with $\alpha_2 = 0.4$. 
In order to find a feasible waveform and make sure ${\cal{P}}_r^S$ is always feasible, we first solve the following optimization problem

\begin{equation}
\begin{array}{l}
{{\bm{s}}^{(0)}} = {\kern 1pt} \arg \mathop {\min }\limits_{\bm{s}} {{\bm{s}}^{\rm{H}}}{{\bm{R}}_I}{\bm{s}}\\
s.t.\;\;\;{\kern 1pt} \left\| {\bm{s}} \right\|_2^2 \le {e_t},{\left\| {{\bm{s}} - {{\bm{s}}_0}} \right\|_\infty } \le \frac{{\delta \sqrt {{e_t}} }}{{\sqrt {{N_T}L} }}
\end{array},
\end{equation}
then set ${e_I} \ge {\kern 1pt} {\left( {{{\bm{s}}^{(0)}}} \right)^{\rm{H}}}{{\bm{R}}_I}{{\bm{s}}^{(0)}}$. Given the fact that there is no available algorithm solving the minimax waveform-filter design problem for extended target under SC-SC, we only carried out some experiments to test \textbf{Algorithm 3} in this subsection. 

Now, we turn to the convergence of \textbf{Algorithm 3}. Fig.{\ref{IterationCurveCMSC}} depicts the iteration curve of ${{\tilde z}^S}({\bm{s}^{(l)}},{{\bm{s}}^{(l-1)}})$ with fixed $e_t = 20$dB and $\delta = 1$. Obviously, \textbf{Algorithm 3} converges in several iterations, and SINR monotonically increases with respect to $e_I$. Furthermore, Fig.\ref{WaveformPropertiesSCSC} shows some properties of the transmit waveform. In particular, Fig.\ref{WaveformPropertiesSCSC}(a) depicts the PSD of transmit waveform. One can see from Fig.\ref{WaveformPropertiesSCSC}(a) that the transmit waveform forms deep notches in the limited frequency bands, and the smaller $\delta$ results in deeper notches. While, the corresponding pulse compression results are given in Fig.\ref{WaveformPropertiesSCSC}(b). It is interesting that the parameter $e_I$ also affects the sidelobe level under the same $\delta$. These phenomena are consistent with the results in \cite{aldayel2016successive,aubry2014radar}.

Finally, we randomly pick up some samples from $\Omega$ and calculate the actually output SINR to verify that the designed waveform-filter pair is the Stackelberg equilibrium strategy for radar. The corresponding results are shown in Fig.\ref{VerifyRobust}, and we are glad to see that the actually achieved SINR is significantly higher than the SINR(worst-case)  optimized by \textbf{Algorithm 3}, which demonstrates the effectiveness of our algorithm.  

\begin{figure}[!t]
	\centering
	\includegraphics[width=3.2in]{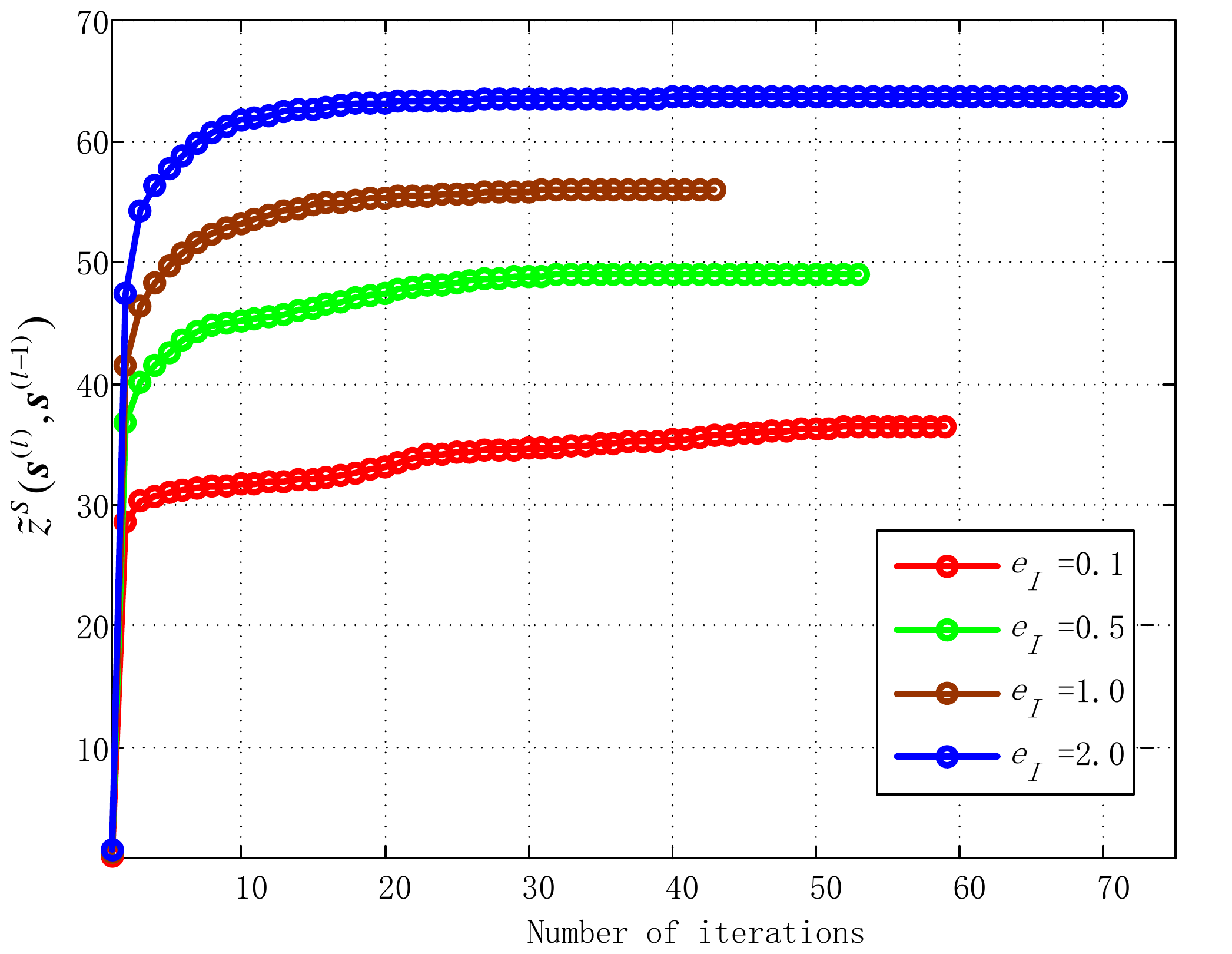}
	\caption{${{\tilde z}^S}({\bm{s}^{(l)}},{{\bm{s}}^{(l-1)}})$ versus the number of iterations for different $e_I$}
	\label{IterationCurveCMSC}
\end{figure}

\begin{figure}[!t]
	\centering
	\begin{tabular}{c}
		\includegraphics[width=3.0in]{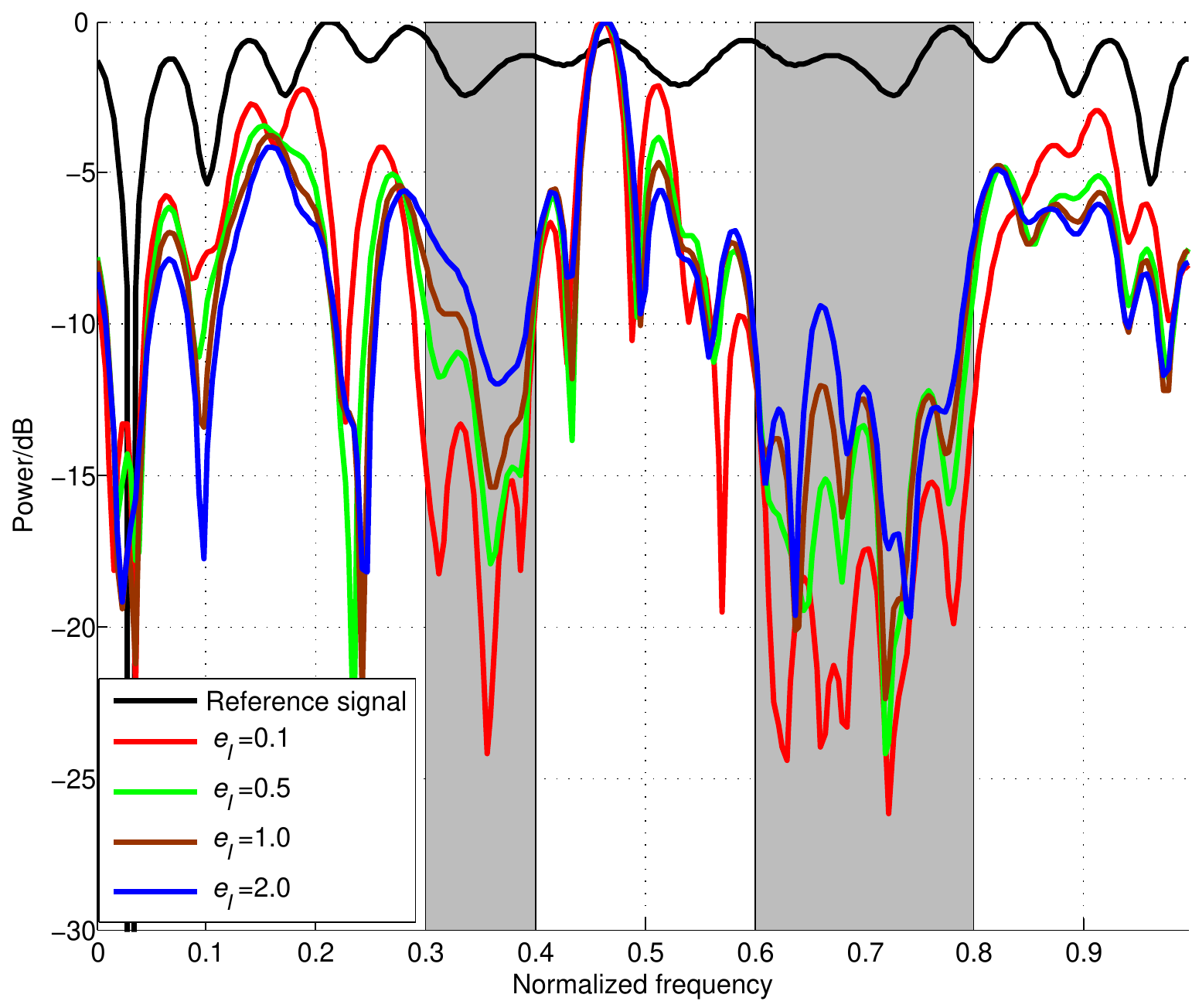}\\
		(a)\\	
		\includegraphics[width=3.0in]{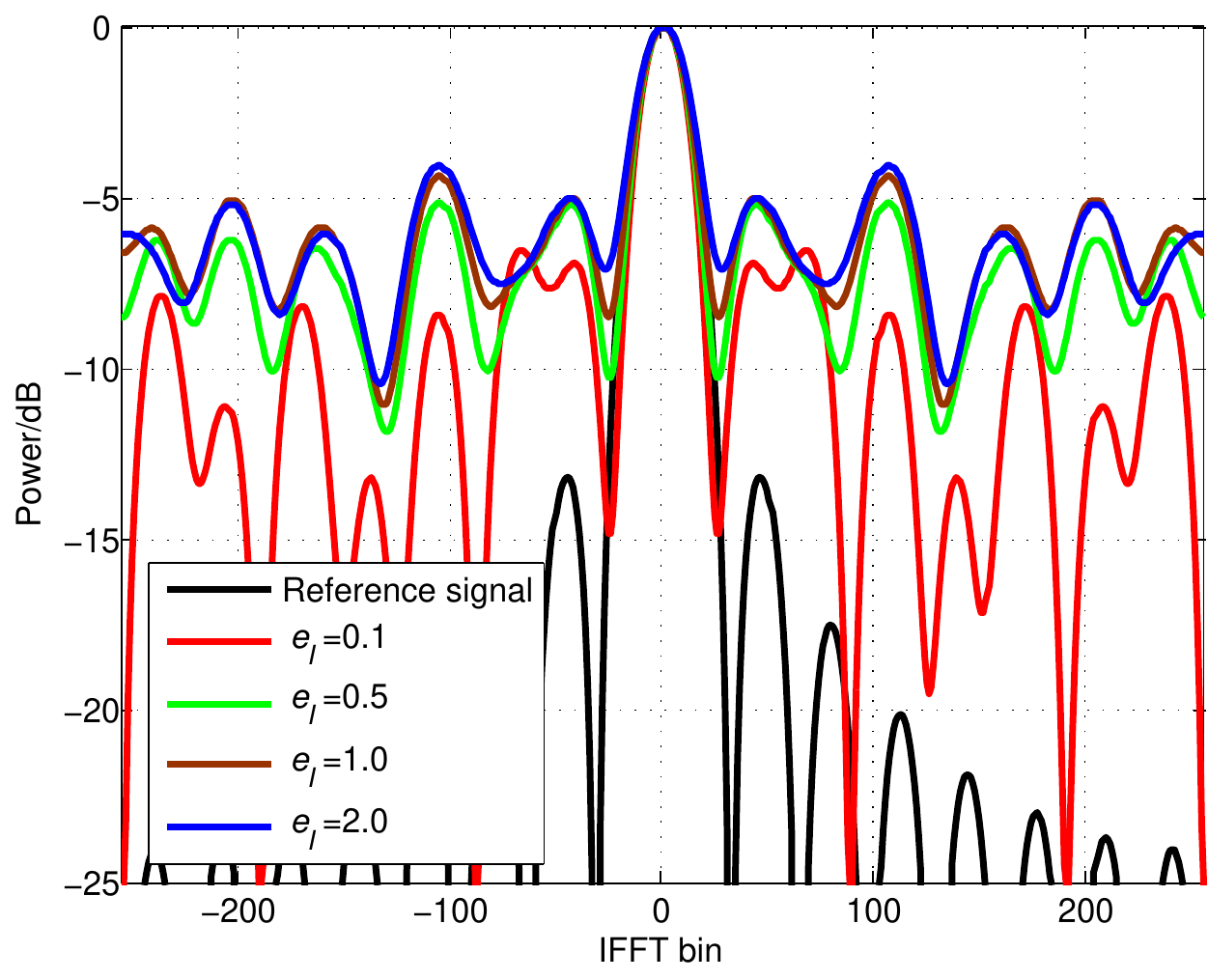}\\
		(b)
	\end{tabular}
	\centering
	\caption{Waveform properties for different $e_I$: (a) Power Spectral Density,  (b) Pulse Compression}
	\label{WaveformPropertiesSCSC}
\end{figure}

\begin{figure}[!t]
	\centering
	\includegraphics[width=3.2in]{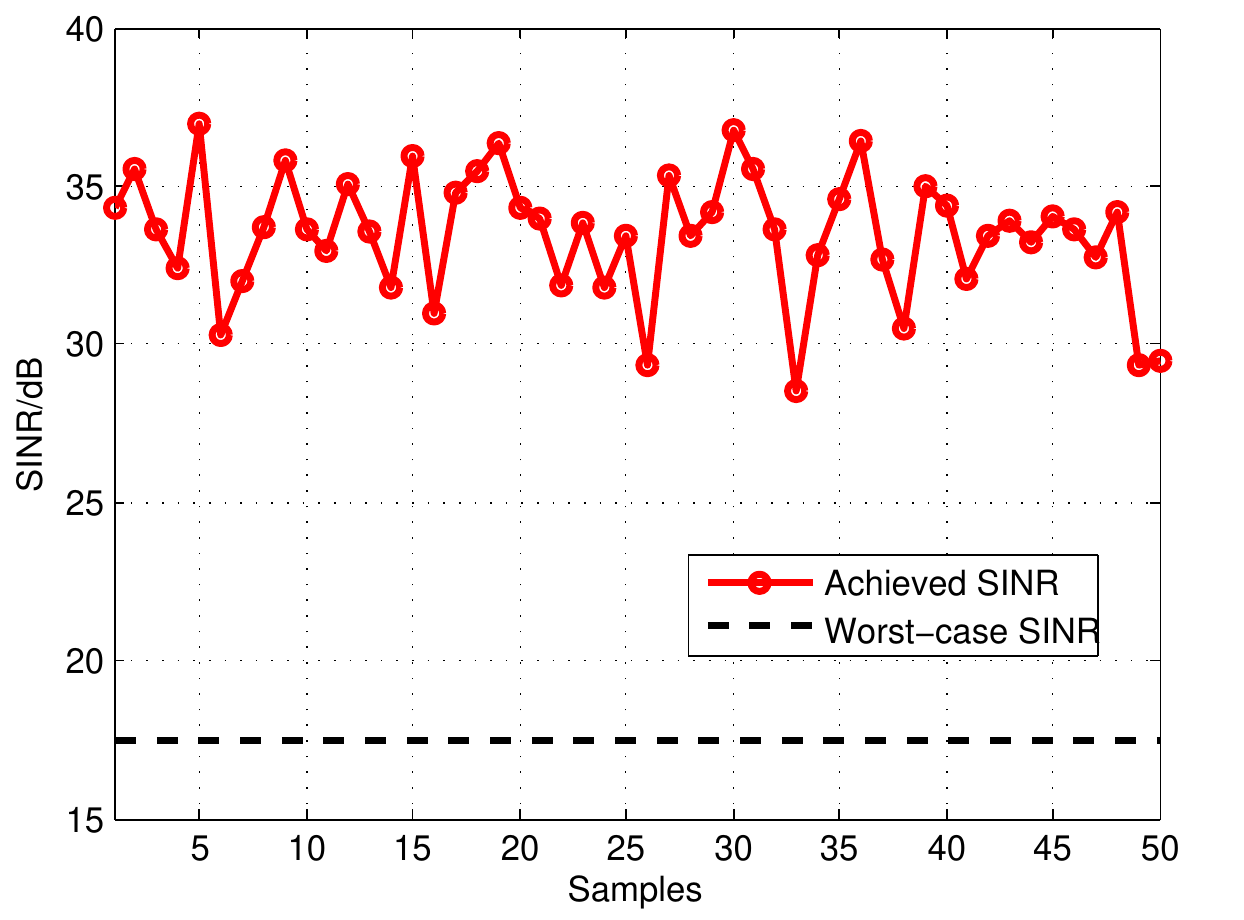}
	\caption{Actually achieved SINR with random samples}
	\label{VerifyRobust}
\end{figure}

\section{Conclusion}
In this paper, we study the joint design problem of transmit waveform and receive filter for extended target detection from the view of TPZS games, where the SINR is used as the payoff function. We assume that the radar player aims at maximizing SINR by choosing its waveform and filter from the strategy set, while the target player is smart enough to minimize SINR by changing its TIR from the strategy set to prevent being detected. The interaction between the radar and target is modeled as a Stackelberg game where radar acts as the leader, and the goal is to find the optimal strategy of radar from Stackelberg equilibrium. The strategy set of target is constrained in a scaled sphere centered by a prescribed TIR. As to the strategy set of radar, the following three cases are studied: 1) EC on Waveform; 2) CM-SC on waveform; 3) SC-SC on waveform. We resort to \textbf{Algorithm 1}, \textbf{Algorithm 2} and \textbf{Algorithm 3} to address the three cases mentioned above, respectively. The conclusions are drawn as follows:

1) Under the EC on waveform, the Stackelberg equilibrium is also the Nash equilibrium, which means that the play who acts as the leader in the game makes no difference on the equilibrium strategies. Thus, in \textbf{Algorithm 1}, the equilibrium strategy is be solved from the game where target acts as the leader, which can be convert to a convex optimization problem and solved in polynomial time. \textbf{Algorithm 1} achieves the same performance as the algorithm in \cite{chen2009mimo} in terms of detection probability, but shows superiority on running time.
  
2) When considering CM-SC on waveform, the Stackelberg equilibrium may not be the Nash equilibrium, but the Nash equilibrium can be approximately constructed from its relaxation form by optimizing the covariance of transmit waveform. \textbf{Algorithm 2} focuses on solving the Nash equilibrium of the relaxation form, and synthesize the Stackelberg equilibrium strategy of the original game with randomization process. Even though it is time consuming compared with the algorithm in \cite{karbasi2015robust}, it achieves higher detection probability. Moreover, the tighter similarity constraint on waveform results in better pulse compression properties but  lower detection probability.

3) As to SC-SC on waveform, the Stackelberg equilibrium are solved leveraging on the MM algorithm, which is the core idea of \textbf{Algorithm 3}. Accordingly, the optimal waveform forms deep notches in the limited frequency bands and show relatively satisfactory pulse compression properties. Interestingly, the spectral compatibility parameter not only affects the its PSD, but also its pulse compression properties. Thus, a suitable trade-off among the output SINR, the similarity level and the spectral compatibility property should be made in practice. Moreover, the Stackelberg equilibrium strategy for radar is essentially the robust waveform-filter.      

Our future researches may include the equilibrium strategies in the presence of signal dependent clutter. Besides, researches on  more practical strategy set of radar and target will be also interesting.

\appendices
\section{Proof of Theorem 1}

For a fixed $\bm{s}^0$, the optimization problem in ${\tilde{\cal P}_r}$ can be recast as 
\begin{equation}
F(\bm{s}^0)= \mathop {{\rm{min}}}\limits_{{\bm{t}} \in \Omega } \mathop {{\rm{max}}}\limits_{\bm{w}} {\rm{ }}\frac{{{{\bm{w}}^{\rm{H}}}{\bm{y}}_t^0{{\left( {{\bm{y}}_t^0} \right)}^{\rm{H}}}{\bm{w}}}}{{{{\bm{w}}^{\rm{H}}}{{\bm{R}}_c}{\bf{w}}}} 
\end{equation}
where ${{\bm{y}}_t^0} = {\bm{H}}({\bm{s}^0}){\bm{t}}$. On the other hand, the optimization problem in ${{\cal P}_r}$ can be recast as 
\begin{equation}
G(\bm{s}^0) = \mathop {{\rm{max}}}\limits_{\bm{w}} \mathop {{\rm{min}}}\limits_{{\bm{t}} \in \Omega } {\rm{ }}\frac{{{{\bm{w}}^{\rm{H}}}{\bm{y}}_t^0{{\left( {{\bm{y}}_t^0} \right)}^{\rm{H}}}{\bm{w}}}}{{{{\bm{w}}^{\rm{H}}}{{\bm{R}}_c}{\bf{w}}}}.
\end{equation}

Further, by solving the inner maximization problem of $F(\bm{s}^0)$, we have  
\begin{equation}
F({{\bm{s}}^0}) = \mathop {{\rm{min}}}\limits_{{\bm{t}} \in \Omega } {\rm{ }}{\left( {{\bm{y}}_t^0} \right)^{\rm{H}}}{\bm{R}}_c^{ - 1}{\bm{y}}_t^0 = \mathop {{\rm{min}}}\limits_{{\bm{t}} \in \Omega } {\rm{ }}{{\bm{t}}^{\rm{H}}}{\left( {{\bm{H}}({{\bm{s}}^0})} \right)^{\rm{H}}}{\bm{R}}_c^{ - 1}{\bm{H}}({{\bm{s}}^0}){\bm{t}}
\label{MinTFixedS0}
\end{equation}
Note that $F(\bm{s}^0) \geq G(\bm{s}^0)$ always holds due to the weak minimax inequality, which means 
\begin{equation}
G(\bm{s}^0) \le \mathop {{\rm{min}}}\limits_{{\bm{t}} \in \Omega } {\rm{ }}{{\bm{t}}^{\rm{H}}}{\left( {{\bm{H}}({{\bm{s}}^0})} \right)^{\rm{H}}}{\bm{R}}_c^{ - 1}{\bm{H}}({{\bm{s}}^0}){\bm{t}}..
\label{WeakMinimaxW}
\end{equation}

Denote by $\bm{t}_{s_0}^*$ the optimal solution of (\ref{MinTFixedS0}), and let $\bm{w}_{s_0}^*={\bm{R}}_c^{ - 1}{\bm{H}}({{\bm{s}}^0}){\bm{t}_{s_0}^*}$, then we have 
\begin{equation}
G({{\bm{s}}^0}) \ge \mathop {{\rm{max}}}\limits_{{\bm{w}}=\bm{w}_{s_0}^*} \mathop {{\rm{min}}}\limits_{{\bm{t}} \in \Omega } {\rm{ }}\frac{{{{\bm{w}}^{\rm{H}}}{\bm{y}}_t^0{{\left( {{\bm{y}}_t^0} \right)}^{\rm{H}}}{\bm{w}}}}{{{{\bm{w}}^{\rm{H}}}{{\bm{R}}_c}{\bm{w}}}}.
\label{StrongMinimaxW}
\end{equation}
The inequality holds due to the shrinkage of feasible set on $\bm{w}$.

The right hand in (\ref{StrongMinimaxW}) is equal to $F({{\bm{s}}^0})$ according to Theorem 1 in \cite{kim2006robust} given that $\Omega$ is convex.

Combining (\ref{WeakMinimaxW}) and (\ref{StrongMinimaxW}), one can obtain the following strong minimax inequality
\begin{equation}
\mathop {{\rm{max}}}\limits_{\bm{w}} \mathop {{\rm{min}}}\limits_{{\bm{t}} \in \Omega } {\rm{ }}\frac{{{{\bm{w}}^{\rm{H}}}{\bm{y}}_t^0{{\left( {{\bm{y}}_t^0} \right)}^{\rm{H}}}{\bm{w}}}}{{{{\bm{w}}^{\rm{H}}}{{\bm{R}}_c}{\bm{w}}}} = \mathop {{\rm{min}}}\limits_{{\bm{t}} \in \Omega } \mathop {{\rm{max}}}\limits_{\bm{w}} {\rm{ }}\frac{{{{\bm{w}}^{\rm{H}}}{\bm{y}}_t^0{{\left( {{\bm{y}}_t^0} \right)}^{\rm{H}}}{\bm{w}}}}{{{{\bm{w}}^{\rm{H}}}{{\bm{R}}_c}{\bm{w}}}}.
\label{StrongMinimaxWEquality}
\end{equation}

Recall that (\ref{StrongMinimaxWEquality}) holds for any $\bm{s}^0$, then we have 
\begin{equation}
\mathop {{\rm{max}}}\limits_{{\bm{s}} \in \Psi ,{\bm{w}}} \mathop {{\rm{min}}}\limits_{{\bm{t}} \in \Omega } {\rm{ }}\frac{{{{\bm{w}}^{\rm{H}}}{{\bm{y}}_t}{\bm{y}}_t^{\rm{H}}{\bm{w}}}}{{{{\bm{w}}^{\rm{H}}}{{\bm{R}}_c}{\bm{w}}}} =\mathop {{\rm{max}}}\limits_{{\bm{s}} \in \Psi } \mathop {{\rm{min}}}\limits_{{\bm{t}} \in \Omega } \mathop {{\rm{max}}}\limits_{{\bm{w}}} {\rm{ }}\frac{{{{\bm{w}}^{\rm{H}}}{{\bm{y}}_t}{\bm{y}}_t^{\rm{H}}{\bm{w}}}}{{{{\bm{w}}^{\rm{H}}}{{\bm{R}}_c}{\bm{w}}}},
\end{equation}
and the optimal solutions for ${\cal{P}}_r$ and $\tilde {\cal{P}}_r$ are the same. 
 
Thus, we complete the proof of Theorem 1.

\section{Proof of Proposition 1}
The proof starts with $\tilde {\cal P}_t^E$ described in (\ref{OptimizationLambda}), which is equivalent to the following problem 
\begin{equation}
\left\{ {\begin{array}{*{20}{l}}
	{{{\overline {{\rm{SINR}}} }_E} = \mathop {{\rm{min}}}\limits_{\bm{G(t)}} {\lambda _{\max }}\left( {{\bm{G}}{{({\bm{t}})}^{\rm{H}}}{\bm{R}}_c^{ - 1}{\bm{G}}({\bm{t}})} \right){e_t}}\\
	{s.t.\;\;\;{\kern 1pt} {\bm{G}}({\bm{t}}) \in \tilde \Omega }
	\end{array}} \right.,
\label{OptimizationGt}
\end{equation}
where $\tilde \Omega  = \left\{ {{\bm{G}}({\bm{t}})|\bm{t} \in \Omega} \right\}$ is also convex since $\bm{G(t)}$ is an affine function with respect to $\bm{t}$.

Denote by $\bm{t}_E^*$ the optimal solution of $\tilde {\cal P}_t^E$, then $\bm{G}\left( \bm{t}_E^*\right) $ is an optimal solution of (\ref{OptimizationGt}), which meets the following optimality condition 
\begin{equation}
\begin{array}{l}
\left\langle {{{\left. {{\nabla _{{\bm{G}}({\bm{t}})}}{\lambda _{\max }}\left( {{\bm{G}}{{({\bm{t}})}^{\rm{H}}}{\bm{R}}_c^{ - 1}{\bm{G}}({\bm{t}})} \right)} \right|}_{{\bm{G}}({\bm{t}}_E^*)}},{\bm{G}}({\bm{t}}) - {\bm{G}}({\bm{t}}_E^*)} \right\rangle \\
\ge 0,{\rm{ }}\forall {\bm{G}}({\bm{t}}) \in \tilde \Omega
\end{array}.
\label{OptimalityCondition}
\end{equation}
After some algebraic manipulations \cite{hjorungnes2011complex}, (\ref{OptimalityCondition}) reduces to 
\begin{equation}
{\left( {{\bm{s}}_E^*} \right)^{\rm{H}}}{\bm{G}}{({\bm{t}}_E^*)^{\rm{H}}}{\bm{R}}_c^{ - 1}\left( {{\bm{G}}({\bm{t}}) - {\bm{G}}({\bm{t}}_E^*)} \right){\bm{s}}_E^* \ge 0,\forall {\bm{G}}({\bm{t}}) \in \tilde \Omega 
\end{equation}
with ${{\bm{s}}_E^*}$ defined in (\ref{OptimalWaveformEigenVector}).

On the other hand, 
\begin{equation}
\begin{array}{l}
{\underline {{\rm{SINR}}} _E} = \mathop {{\rm{max}}}\limits_{{\bm{s}} \in {\Psi _E}} \mathop {{\rm{min}}}\limits_{{\bm{t}} \in \Omega } {{\bm{t}}^{\rm{H}}}{\bm{H}}{({\bm{s}})^{\rm{H}}}{\bm{R}}_c^{ - 1}{\bm{H}}({\bm{s}}){\bm{t}}\\
\quad \quad \quad  \; \; \ge \mathop {{\rm{min}}}\limits_{{\bm{t}} \in \Omega } {{\bm{t}}^{\rm{H}}}{\bm{H}}{({\bm{s}}_E^*)^{\rm{H}}}{\bm{R}}_c^{ - 1}{\bm{H}}({\bm{s}}_E^*){\bm{t}}
\end{array}.
\label{StrongMinimax1}
\end{equation}
Next, we are going to prove that the value of right hand in inequality (\ref{StrongMinimax1}) is \vspace{2pt}equal to ${\overline {{\rm{SINR}}} _E}$, i.e., $\bm{t}_E^*$ meets the optimality condition of \vspace{2pt} $\mathop {{\rm{min}}}\limits_{{\bm{t}} \in \Omega } {{\bm{t}}^{\rm{H}}}{\bm{H}}{({\bm{s}}_E^*)^{\rm{H}}}{\bm{R}}_c^{ - 1}{\bm{H}}({\bm{s}}_E^*){\bm{t}}$.

Denote by $\tilde{\bm{t}}_E^*$ the optimal solution of $\mathop {{\rm{min}}}\limits_{{\bm{t}} \in \Omega } {{\bm{t}}^{\rm{H}}}{\bm{H}}{({\bm{s}}_E^*)^{\rm{H}}}{\bm{R}}_c^{ - 1}{\bm{H}}({\bm{s}}_E^*){\bm{t}}$, then we get the following optimality condition 
\begin{equation}
{\left( {{\bm{t}} - {\bm{\tilde t}}_E^*} \right)^{\rm{H}}}{\bm{H}}{({\bm{s}}_E^*)^{\rm{H}}}{\bm{R}}_c^{ - 1}{\bm{H}}({\bm{s}}_E^*){\bm{\tilde t}}_E^* \ge 0,\forall {\bm{t}} \in \Omega.
\label{StrongMinimax2}
\end{equation}
Note that $\bm{G(t)}$ is the one-to-one mapping with respect to $\bm{t}$. Therefore, combining (\ref{StrongMinimax1}) and $\bm{G}{(\bm{t})\bm{s}} = \bm{H}{(\bm{s})\bm{t}}$, we get that $\bm{t}_E^*$ also meets the optimality condition (\ref{StrongMinimax2}), namely,
\begin{equation}
\underline {\rm{SINR}}_E \ge \overline {\rm{SINR}}_E.
\end{equation}

Furthermore, the weak minimax inequality \vspace{2pt} $\underline {\rm{SINR}}_E \le \overline {\rm{SINR}}_E$ always holds. To this end, we have 
\begin{equation}
\underline {\rm{SINR}}_E = \overline {\rm{SINR}}_E,
\end{equation}
and the strategy pair $(\bm{s}_E^*,{\bm{R}}_c^{ - 1}{\bm{G}}({\bm{t}}_E^*){\bm{s}}_E^*,\bm{t}_E^*)$ is Nash equilibrium.

Thus we complete the proof of Proposition 1.

\section{Proof of Proposition 2}
For any $\bm{t}$, we have 
\begin{equation}
\begin{array}{l}
{{\bm{s}}^{\rm{H}}}{\bm{G}}{({\bm{t}})^{\rm{H}}}{\bm{R}}_c^{ - 1}{\bm{G}}({\bm{t}}){\bm{s}} = 2{\mathop{\rm Re}\nolimits} \left\{ {{{\left( {{{\bm{s}}^{(l)}}} \right)}^{\rm{H}}}{\bm{G}}{{({\bm{t}})}^{\rm{H}}}{\bm{R}}_c^{ - 1}{\bm{G}}({\bm{t}}){\bm{s}}} \right\} - \\
\quad \quad \quad \quad \quad \quad \quad \quad {\left( {{{\bm{s}}^{(l)}}} \right)^{\rm{H}}}{\bm{G}}{({\bm{t}})^{\rm{H}}}{\bm{R}}_c^{ - 1}{\bm{G}}({\bm{t}}){{\bm{s}}^{(l)}} + \\
\quad \quad \quad \quad \quad \quad \quad \quad {\left( {{\bm{s}} - {{\bm{s}}^{(l)}}} \right)^{\rm{H}}}{\bm{G}}{({\bm{t}})^{\rm{H}}}{\bm{R}}_c^{ - 1}{\bm{G}}({\bm{t}})\left( {{\bm{s}} - {{\bm{s}}^{(l)}}} \right)\\
\end{array}.
\end{equation}
Given the fact that ${\bm{G}}{({\bm{t}})^{\rm{H}}}{\bm{R}}_c^{ - 1}{\bm{G}}({\bm{t}}) \succeq 0$, we get the following inequality
\begin{equation}
\begin{array}{l}
{{\bm{s}}^{\rm{H}}}{\bm{G}}{({\bm{t}})^{\rm{H}}}{\bm{R}}_c^{ - 1}{\bm{G}}({\bm{t}}){\bm{s}} \ge 2{\mathop{\rm Re}\nolimits} \left\{ {{{\left( {{{\bm{s}}^{(l)}}} \right)}^{\rm{H}}}{\bm{G}}{{({\bm{t}})}^{\rm{H}}}{\bm{R}}_c^{ - 1}{\bm{G}}({\bm{t}}){\bm{s}}} \right\} - \\
\quad \quad \quad \quad \quad \quad \quad \quad {\left( {{{\bm{s}}^{(l)}}} \right)^{\rm{H}}}{\bm{G}}{({\bm{t}})^{\rm{H}}}{\bm{R}}_c^{ - 1}{\bm{G}}({\bm{t}}){{\bm{s}}^{(l)}}
\end{array},
\end{equation}
with equality if and only if $\bm{s}=\bm{s}^{(l)}$.

Consequently, ${{\tilde z}^S}({\bm{s}},{{\bm{s}}^{(l)}})$ 
is a minorizer of ${z^S}({\bm{s}})$ at $\bm{s}^{(l)}$.

Thus we complete the proof of Proposition 2.

\section{Proof of Proposition 3}
Let us start with
\begin{equation*}
{\tilde z^S}({\bm{s}},{{\bm{s}}^{(l)}}) = \mathop {{\rm{min}}}\limits_{\left\{ {{\bm{t}}|{{\left\| {{\bm{t}} - {{\bm{t}}_0}} \right\|}_2} \le r} \right\}} {{\bm{t}}^{\rm{H}}}{\bm{U}}({\bm{s}},{{\bm{s}}^{(l)}}){\bm{t}}
\end{equation*}
described in ({\ref{ZSt}}). And the dual problem is given by 
\begin{equation}
\begin{array}{l}
\mathop {\max }\limits_{\lambda ,\gamma } \;\gamma \\
s.t.\left[ {\begin{array}{*{20}{c}}
	{{\bm{U}}({\bm{s}},{{\bm{s}}^{(l)}}) + \lambda {\bf{I}}}&{\lambda {{\bm{t}}_0}}\\
	{\lambda {\bm{t}}_0^{\rm{H}}}&{\lambda {\bm{t}}_0^{\rm{H}}{{\bm{t}}_0} - \lambda {r^2} - \gamma }
	\end{array}} \right] \succeq 0\quad \\
\quad \lambda  \ge 0
\end{array}.
\end{equation}
Note that it is a single constraint quadratic problem and strong duality holds, even though it is non-convex{\cite{BoydConvex}}. Therefore, we reformulate it as the dual form and get $\hat{{\cal{P}}}_r^{S,(l)}$ by substituting the dual form into ${\cal{P}}_r^{S,(l)}$.

Thus we complete the proof of Proposition 3.

\section*{Acknowledgment}

\bibliographystyle{IEEEtran}
\bibliography{MIMORadarWaveformFilterDesignforExtendedTargetDetectionfromaViewofGames}     

\end{document}